\documentclass[12pt,draftclsnofoot,onecolumn,romanappendices]{IEEEtran}

\makeatletter
\def\ps@headings{%
\def\@oddhead{\mbox{}\scriptsize\rightmark \hfil \thepage}%
\def\@evenhead{\scriptsize\thepage \hfil\leftmark\mbox{}}%
\def\@oddfoot{}%
\def\@evenfoot{}}
\makeatother
\usepackage{latexsym}
\usepackage{amssymb}
\usepackage{epsfig}
\usepackage{amsthm} 
\usepackage{subfigure}
\usepackage{graphics,color}
\usepackage{graphicx}
\usepackage{amsmath}
\usepackage[ruled,vlined,boxed,linesnumbered]{algorithm2e}
\usepackage{cite}
\usepackage{comment}
\usepackage{url}
\usepackage{xspace}
\usepackage{psfrag}
\usepackage{booktabs}
\usepackage{stfloats}
\usepackage{balance}

\newtheorem{theorem}{Theorem}
\newtheorem{lemma}{Lemma}
\newtheorem{proposition}{Proposition}
\newtheorem{corollary}{Corollary}
\newtheorem{assumption}{Assumption}
\newtheorem{fact}{Fact}
\theoremstyle{definition}
\newtheorem{definition} {Definition}
\newtheorem*{remarks}{Remark}

\newcommand{\ignore}[1]{}

\newcommand{\tj}[1]{{\color{blue}{#1}}}

\def\calA{\mathcal{A}}
\def\calH{\mathcal{H}}
\def\calX{\mathcal{X}}
\def\calY{\mathcal{Y}}
\def\calL{\mathcal{L}}

\def\fc{{\mathfrak c}}
\def\Pe{\mathrm {Pe}}

\def\tmin{\mathrm {min}}

\def\taue{{\tau_\epsilon}}
\def\tause{{\tau^*_\epsilon}}
\def\taute{{\tilde\tau_\epsilon}}
\def\tauti{{\tilde\tau_\iota^*}}
\def\tauce{{\tau^{\mathfrak c}_\epsilon}}

\DeclareMathOperator*{\argmax}{arg\,max}

\newcommand{\changetokc}[2]{#2}

\begin{document}

\title{Noisy Bayesian Active Learning}
\author{Mohammad Naghshvar, Tara Javidi, and Kamalika Chaudhuri 
\thanks{This paper was presented in part at Allerton 2012 and 2013. 

This work was done while M. Naghshvar was with the Department of Electrical and Computer Engineering, University of California San Diego, La Jolla, CA 92093 USA.
He is now with Qualcomm Technologies Inc., San Diego, CA 92121 USA (e-mail: mnaghshvar@qti.qualcomm.com).
T. Javidi is with the Department of Electrical and Computer Engineering, University of California San Diego, La Jolla, CA 92093 USA (e-mail: tjavidi@ucsd.edu).
K. Chaudhuri is with the Department of Computer Science, University of California San Diego, La Jolla, CA 92093 USA. (e-mail: kchaudhuri@ucsd.edu).

The work of M.~Naghshvar and T.~Javidi was partially supported by the industrial sponsors of 
UCSD Center for Wireless Communication (CWC), Information Theory and Applications Center (ITA), 
and Center for Networked Systems (CNS), as well as NSF Grants CCF-0729060 and CCF-1018722. 
The work of K.~Chaudhuri was partially supported by NSF under IIS-1162581.
}}

\maketitle

\thispagestyle{empty}

\begin{abstract}
We consider the problem of noisy Bayesian active learning, where we are given a finite set of functions $\mathcal{H}$, a sample space $\mathcal{X}$, and a label set $\mathcal{L}$. One of the  functions in $\mathcal{H}$ assigns labels to samples in $\mathcal{X}$. 
The goal is to identify the function that generates the labels even though the result of a label query on a sample is corrupted by independent noise. More precisely, the objective is to declare one of the functions in $\mathcal{H}$ as the true label generating function  
 with high confidence using as few label queries as possible, by selecting the queries adaptively and in a strategic manner.  

Previous work in Bayesian active learning considers Generalized Binary Search, and its variants for the noisy case, and analyzes the number of queries required by these sampling strategies.  In this paper, we show that these schemes are, in general, suboptimal.  Instead we propose and analyze an alternative strategy for sample collection.  Our sampling strategy is motivated by a connection between Bayesian active learning and active hypothesis testing, and is based on querying the label of a sample which maximizes the Extrinsic Jensen--Shannon divergence at each step. 
We provide upper and lower bounds on the performance of this sampling strategy, and show that these bounds are better than previous bounds.
\end{abstract}


{\allowdisplaybreaks[4]{
\section{Introduction}


We consider the problem of noisy Bayesian active learning, where we are given a finite set of functions $\mathcal{H}$, a sample space $\mathcal{X}$, and a label set $\mathcal{L}$. One of the  functions in $\mathcal{H}$ assigns labels to samples in $\mathcal{X}$, and our goal is to identify this function when the result of a label query on a sample is corrupted by independent noise. The objective is to declare one of the functions in $\mathcal{H}$ as the true label generating function with high confidence using as few label queries as possible, by selecting the queries adaptively and in a strategic manner.

A special case of the problem, first considered by \cite{Nowak09},  arises when the label set is binary and the natural
sampling strategy for Bayesian active learning becomes closely related to Generalized Binary Search (GBS). In the binary label setting, GBS queries the label of a sample $x$ for which the size of the subsets of functions that label $x$ as $+1$ and $-1$ respectively, are as balanced as possible. A variant of GBS is Modified Soft-Decision Generalized Binary Search (MSGBS), which was introduced by \cite{Nowak09} to address the case when the observed labels may be noisy. \cite{Nowak09} analyzes the performance of MSGBS, under a symmetric and non-persistent noise model which flips the labels randomly, and shows that the \changetokc{(fixed)}{} number of samples required to identify the correct function with probability of error satisfying $\text{Pe}\le\epsilon$ is $O\left( \frac{\log M+\log\frac{1}{\epsilon}}{\lambda} \right)$, where $M$ is the number of functions in the class $\mathcal{H}$, and $\lambda$ is a parameter which depends on the structure of the function class, the sample space, and the noise rate. 
The \changetokc{main}{first} contribution of this paper is to generalize the above problem to the case of general (non-binary) label set with general (and potentially non-symmetric) non-persistent observation noise.

By allowing for the number of samples collected to be determined in a sequential manner (according to a random stopping time as a function of past observations), we draw a parallel between active sequential hypothesis testing and Bayesian active learning. 
In active sequential hypothesis testing, we are given a set of $M$ hypotheses, and a set of actions; each action, conditioned 
on the true hypothesis, has a certain probability of yielding an outcome. We observe that Bayesian active learning is a special case of active hypothesis testing, where the hypotheses map to functions, actions map to samples, and the outcomes map to noisy observation of labels. This view of the problem allows for a natural extension of the model of~\cite{Nowak09} to the non-binary Bayesian active learning setting, where the label noise might be label dependent and asymmetric.  Relying on this connection, we
derive a universal lower bound on the \emph{expected} number of samples required to identify the true hypothesis among $M$ with reliability $\epsilon$ as a function of noise model parameters. 
Our lower bound generalizes that of~\cite{Burnashev76}. 
This lower bound, when specialized for the noisy generalized binary search suggests that the
proposed schemes of  \cite{Nowak09} are suboptimal in general. The next contribution of this work is to propose and analyze an alternative strategy for sample collection.

To find an alternative strategy, we again take advantage of the connection between Bayesian learning and active sequential hypothesis testing. In~\cite{ISIT2012}, the authors introduced the notion of Extrinsic Jensen--Shannon (EJS) divergence, 
and proposed an active sequential hypothesis test that, at each step, selects the action that maximizes the EJS divergence.
%
%
%
In this paper, we apply the corresponding sampling strategy to Bayesian active learning, and characterize the performance of this strategy. Our analysis improves on the analysis of~\cite{ISIT2012}. Our bounds show that the number of label queries required by our algorithm is $O\left(\frac{\log M}{\alpha}+\frac{\log\frac{1}{\epsilon}}{\beta}\right)$, where $M$ is the number of functions and $\alpha$ and $\beta$ are terms, different from $\lambda$, that depend on the structure of the function class, the sample space, and the noise model.
  
To illustrate our bounds, in Section~\ref{SC}, we focus on generalized binary search studied in \cite{Nowak09} and consider the class of 1-neighborly functions and its three specific subclasses --- intervals on the line, thresholds on the line, and a set of rich function classes. 
We show that the upper bounds on the number of labels required by the EJS policy are superior to those of~\cite{Nowak09} for all three subclasses for the asymptotic values of $\epsilon$ and $M$.
%
In addition, we show through numerical simulations that our policy has better performance than the algorithms of~\cite{Nowak09} also in non-asymptotic regimes of practical interest. 

There has been a large explosion of recent work on the theory of active learning~\cite{Dassurvey11, DAS05, Kar06, H07, BBL09, DHM07, BDL09, BHLZ10, RR11, GKR10} but despite the similarity of the titles, the models and the assumptions vary drastically with at times contradictory conclusions. Here we attempt to detail specific attributes of these papers and the connection/disconnect between our work and this literature. Early work on active learning~\cite{Dassurvey11, DAS05, H07} considered the realizable case where the binary labels are produced by a function in a given function class and are observed noise-free. Here, the function class is either finite, like our setting, or, unlike our setting, infinite but equipped with a fixed structure, such as the class of thresholds on a line, or the class of linear classifiers.  In contrast with our work, however, the learner is only allowed to query the labels of samples among an unlabeled set of points which are drawn from the unlabeled data distribution. Also unlike ours, the goal here is to find a function which has low \emph{prediction error with respect to the data distribution}.  Thus the challenge is to identify a function in the function class where the disagreement with the true labeling function is less than the required accuracy, and the prediction error occurs due to infiniteness of the function class or due to the indistinguishability of the functions with respect to the data distribution as opposed to noisy observations of the labels. 

Since the realizability assumption can hardly ever be justified in practice, more recent literature~\cite{Kar06, H07, BBL09, BDL09, BHLZ10, RR11} has considered active learning in the non-realizable case. A line of work~\cite{H07, BBL09, DHM07, BDL09, BHLZ10} considers active learning in the agnostic setting, where the binary labels are not necessarily generated by a function in a given function class, and the goal is to find a function in the function class which has low prediction error with respect to the labeled data distribution. Most of this work employs a {\em{disagreement-based}} strategy for label queries; the algorithm maintains a candidate set of functions that is guaranteed to contain the best function in the class with high probability, and queries the label of a sample only when there are two functions in the candidate set that disagree on its label.  An important special case of the non-realizable setting relevant to our work is the bounded rate class noise of \cite{Kar06} in which labels are produced by a member of a given function class but are subjected to an exogenous (and non-persistent) observation noise. In such a setting, \cite{Kar06, Sakakibara91} show that repeat queries can be effectively utilized to mitigate the effect of noise. In \cite{RR11}, the authors perform an information theoretic analysis of active learning in the agnostic setting and provide lower bounds on its sample complexity. 

Finally, \cite{GKR10} considers the same setting as our work. Unlike us, they do not provide absolute upper and lower bounds on the query complexity. Instead, they consider sampling strategies that select the sample that maximizes the information gain based on a certain measure of information, and show that if the measure of information in question is adaptively submodular, then this strategy is competitive with the optimal strategy according to the same information measure. 

										
In summary, our work differs from the previous work on active learning in three important ways. First, we are interested in a generalized learning setup where labels can be non-binary and observation noise can have a general non-symmetric and non-discrete nature. Second, we are interested in a sequential learning setting where the learner is allowed not only  to query individual samples (hence, rendering the data distribution irrelevant), but also to determine the number of queries in an online fashion as a function of observations so far. Third, by considering the simpler setup of a finite function class as well as an exogenous and non-persistent observation noise, we provide sharp lower and upper bounds on the query complexity.
Our lower bound is purely information theoretic and is only a function of the observation noise which is the only inevitable source of inaccuracy in our model. Our upper bound, in contrast, is obtained via the analysis of an achievable scheme and sheds light on how the structure of the function class impacts the overall performance of our proposed scheme.  Perhaps, most significantly, we show that the number of label queries required by the proposed scheme matches the lower bound asymptotically when the function/sample space is sufficiently ``rich.''

The remainder of this paper is organized as follows. 
In Section~\ref{PF}, we formulate the problem of Bayesian active learning.
In Section~\ref{EJS}, we propose our heuristic policy for selecting samples. 
Section~\ref{Main} provides the main results of the paper.
As a special case, noisy generalized binary search is discussed in Section~\ref{SC}
and a comparison to some of the known results is provided.
Finally, we conclude the paper and discuss future work in Section~\ref{Discussion}.

\underline{Notation}:
Let $[x]^+=\max \{x,0\}$. 
For any set $\mathcal{S}$, $\left| \mathcal{S} \right|$ denotes the cardinality of $\mathcal{S}$.
The space of all probability distributions on set $\calA$ is denoted by $\mathbb{P}(\calA)$. 
%
All logarithms are in base 2.
The entropy function on a vector $\boldsymbol{\rho}=[\rho_1,\rho_2,\ldots,\rho_M] \in [0,1]^M$ 
is defined as $H(\boldsymbol{\rho})=\sum_{i=1}^{M} \rho_i \log\frac{1}{\rho_i}$, 
with the convention that $0 \log\frac{1}{0} = 0$. 
Finally, the Kullback--Leibler (KL) divergence between two probability density functions $q(\cdot)$ and $q'(\cdot)$ on space $\mathcal{Y}$ is defined as $D(q\|q')=\int_{\mathcal{Y}} q(y) \log\frac{q(y)}{q'(y)} dy$,
with the convention $0 \log \frac{a}{0}=0$ and $b \log \frac{b}{0}=\infty$ for $a,b\in [0,1]$ with $b\neq 0$.

\section{Bayesian Active Learning}  
\label{PF}

In this section, we provide the mathematical description of the problem of Bayesian active learning. 


\begin{itemize} 
\item[]{\underline{\bf{Problem~(P)}} [Bayesian Active Learning]}\\
In the Bayesian active learning problem, we are given a \emph{sample space}~$\mathcal{X}$, a finite label set~$\mathcal{L}$, and an \emph{observation space}~$\mathcal{Y}$. We are also given a set $\calH = \{ h_1, h_2, \ldots, h_M \}$ of $M$ distinct functions, where each $h_i: \calX \rightarrow \calL$ maps elements in the sample space $\calX$ to the label set $\calL$. We assume that one of the functions in $\calH$, denoted by $h_\theta$, produces the correct labeling on $\calX$.

The decision maker is allowed to {\em{query}} samples from $\calX$. 
Querying a sample $x$ generates an observation in $y \in \calY$
whose distribution is a given function of the true label as determined by the function $h_\theta$. More specifically, if $h_\theta$ is the true underlying function and hence $l = h_\theta(x)$ is the true label of sample $x$, then the result of a query on $x$ is a $\calY$-valued random variable with probability density $f_{l}(\cdot)$. We assume that the observation densities $\{ f_l(\cdot) \}_{l \in \calL}$ are fixed and known, and observations are conditionally independent over time. 

The goal of the decision maker is to determine the identity of the function in $\calH$ that generates the true labels by an adaptive sequential query of a small number of samples. We assume that the decision maker does not have any extra prior knowledge on the identity of the true function; in other words, it begins with a uniform prior over $\calH$. Let~$\tau$ be the stopping time at which the decision maker retires and declares the  label generating function $h_{\hat{\theta}}$. Furthermore, let $\Pe=P(\hat\theta \neq \theta)$ where $\theta$ is the index of the true function.  In Bayesian active learning, the objective is to design a strategy for the decision maker for querying samples in $\calX$ such that, for any given $\epsilon>0$, we have
\begin{align}
			\label{Obj}
			\text{  minimize  } \mathbb{E} \left[ \tau \right] \text{  subject to  } \Pe\le \epsilon.
		\end{align}
Here 
the minimization is taken over the choice of the stopping time $\tau$ and the learning strategy and
the expectation is taken with respect to the observation distribution as well as the Bayesian uniform prior on the true function in $\calH$.
\end{itemize}

Note that Bayesian learning strategy is more than a single sample query but instead is an adaptive and 
sequential rule that dictates the causal choice of (random) sample queries 
depending on the past observations and past queries prior to the stopping time.  
In this paper, we refer to  this adaptive and sequential rule as a query scheme, $\fc$, 
which together with the particular realization of outputs $Y_{\fc}(0),Y_{\fc}(1),\ldots, Y_{\fc}({\tau_{}-2})$, dictates the sample queries 
$X_{\fc}(1),X_{\fc}(2), \ldots, X_{\fc}({\tau-1})$. 

Before we end this section, and in face of the difficulty in fully characterizing the 
optimal learning strategy in general we define weaker notions of optimality.

\subsection{Asymptotic and Order Optimality}
\label{Asym}

\begin{definition}
\label{AsymDef}
Let $\mathbb{E}_{}[\tauce]$ denote the expected number of samples required by query scheme $\fc$ to achieve $\Pe\le\epsilon$.
Furthermore, let $\mathbb{E}[\tause]$ be the minimum expected number of samples required to achieve $\Pe\le\epsilon$, where the minimum is taken over all possible strategies. 
Query scheme $\fc$ is referred to as \emph{asymptotically optimal} in $\epsilon$ (and $M$) if
\begin{align*} 
(\lim_{M \to \infty}) \lim_{\epsilon \to 0} \frac{\mathbb{E}_{}[\tauce] - \mathbb{E}[\tause]}{\mathbb{E}_{}[\tauce]}=0.
\end{align*}
Query scheme $\fc$ is referred to as \emph{order optimal} in $\epsilon$ (and $M$) if
\begin{align*} 
(\lim_{M \to \infty}) \lim_{\epsilon \to 0} \frac{\mathbb{E}_{}[\tauce] - \mathbb{E}[\tause]}{\mathbb{E}_{}[\tauce]}< 1.
\end{align*}
\end{definition}

It is clear from the definitions above that order optimality is weaker than asymptotic optimality.
If a scheme $\fc$ is asymptotically optimal in $\epsilon$ (and $M$),
then $\mathbb{E}_{}[\tauce]$ and $\mathbb{E}[\tause]$ will have the same dominating terms in $\epsilon$ (and $M$);
while order optimality of scheme $\fc$ only implies that
dominating terms in $\mathbb{E}_{}[\tauce]$ and $\mathbb{E}[\tause]$ are similar up to a constant factor.

\section{Preliminaries and Proposed Heuristic}
\label{EJS}

After providing some preliminary results and notations, including the definition of Extrinsic Jensen--Shannon (EJS) divergence, in this section we propose our EJS-based heuristic.


Let $\Omega = \{1, 2, \ldots, M\}$. Recall that $\theta \in \Omega $ is the random variable that indicates the index of the true function and~$\tau$ is the stopping time at which the decision maker retires and guesses the true index.

Casting the problem as a decision theoretic problem
allows for the structural characterization of the information state, also known as sufficient statistics.
{ Let t}he decision maker's posterior belief about each possible function index $i\in\Omega$, updated
after each sample query and observation for $t=0,1,\ldots,\tau-1$, be 
\begin{equation}
\rho_i(t):= P(\{\theta=i\}| X^{t-1}, Y^{t-1}).
\end{equation}
{The decision maker's posteriors about the true label generating function collectively,} 
\begin{equation}
\boldsymbol{\rho}(t):= [\rho_1(t), \rho_2(t), \ldots, \rho_M(t)],
\end{equation}
form a sufficient statistics for our Bayesian decision maker. In other words, the selection of sample query as a function
of this posterior does not incur any loss of optimality \cite{Bertsekas}. 
\ignore{ Let $d_i$, $i \in \Omega$, represent an action under which the decision maker retires and declares the $i^{\text{th}}$ function as the true label generating function. A Markov stationary deterministic policy $\pi$ is defined as a mapping $\pi: \mathbb{P}(\Omega)\rightarrow \mathcal{X} \cup \{d_1,d_2,\ldots,d_M\}$ 
based on which actions $X(t)$, $t=0,1,\ldots,\tau-1$, stopping time~$\tau$ and the subsequent declaration are selected 
(the choice of any of the retire-declare actions marks the stopping time~$\tau$).   Since 
the decision maker's posterior about the true label generating function 
forms a sufficient statistics for our Bayesian decision maker, restricting sample query strategies to 
that of Markov stationary deterministic policies does not incur any loss of optimality.} In particular, the optimal 
decision maker guesses the function with the highest posterior at time $\tau$ to be the label generating function, i.e.,
\begin{align}
\hat{\theta} = \argmax_{i\in \Omega} \rho_i(\tau). \label{ML_decoder}
\end{align}

{We also note that the dynamics of the information state, i.e., the posterior, follows Bayes' rule. But before we 
make this more precise, let us consider an alternative representation of querying a sample $x \in \calX$:

\begin{definition}
A sample $x\in \mathcal{X}$ generates a $|\calL|$-partition $\Xi^x:=\{{H}_l^x\}_{l\in \mathcal{L}}$ of the function class, i.e., if ${H}_l^x = \{h \in \mathcal{H}:  h(x)=l\}$, then $\mathcal{H} = \cup_{l\in\mathcal{L}} {H}_l^x$. 
\end{definition}
This view allows us to characterize the observation density given the belief vector $\boldsymbol{\rho}$ and queried sample  
$x$ as
\begin{align}
f_x^{\boldsymbol{\rho}}(y) := \sum_{i\in\Omega} \rho_i f_{h_i(x)}(y) = \sum_{l\in\calL} f_l(y) \sum_{i: h_i \in \mathcal{H}_l^x} \rho_i.
\end{align}
Therefore, given the belief vector $\boldsymbol{\rho}(t)$, querying sample $x$ and observing (noisy) label $y$
results in a refinement of the posterior according to the Bayes' rule, i.e.,  
\begin{align}
\boldsymbol{\rho}({t+1}) = {\boldsymbol{\Phi}}^x(\boldsymbol{\rho}(t), y)
\end{align}
where 
\begin{align}
\label{PhiDef}
{\boldsymbol{\Phi}}^x(\boldsymbol{\rho}, y):= \left[ \rho_1 \frac{f_{h_1(x)}(y)}{f_x^{\boldsymbol{\rho}}(y)}, \rho_2 \frac{f_{h_2(x)}(y)}{f_x^{\boldsymbol{\rho}}(y)}, \ldots, \rho_M \frac{f_{h_M(x)}(y)}{f_x^{\boldsymbol{\rho}}(y)} \right].
\end{align}

Many of our results in the paper are obtained as a consequence of a connection between Bayesian active learning and the more general problem of Information Acquisition which has been discussed in full generality in \cite{HypJournal}. 
In particular, taking cue from the
seminal work of DeGroot on statistical decision theory \cite{DeGroot70}, and our own prior work on
active hypothesis testing \cite{ISIT2012}, \ignore{
the above stochastic control view of information acquisition has been used in \cite{ISIT2012}, 
to characterize the performance of any given coding scheme using the expected information utility 
 by the observation.  {Information utility, here, 
generalizes the Shannon theoretic notion of mutual information\cite{DeGroot70},  \cite{ISIT2012} in 
terms of the symmetric divergences JS and EJS}.  More specifically,} given a belief vector 
$\boldsymbol{\rho}\in\mathbb{P}(\Omega)$, the expected utility of the sample query 
$x\in\mathcal{X}$,  or equivalently its corresponding $|\calL|$-partition  $\Xi^x=\{{H}_l^x\}_{l\in \mathcal{L}}$,  
can be characterized by its Extrinsic Jensen--Shannon divergence \cite{ISIT2012}:
\begin{align} \label{EJS-x} 
EJS(\boldsymbol{\rho},x)  & : = \sum_{l\in\calL} \sum_{i: h_i \in \mathcal{H}_l^x} 
\rho_i D\left(f_l\  \bigg\|\  \frac{f_x^{\boldsymbol{\rho}} - \rho_i f_l}{1 - \rho_i} \right). 
\end{align}

We use this to construct our proposed heuristic deterministic Markov sample query strategy.

\subsection{Proposed Heuristic}

In this work, we focus on the following (possibly suboptimal) stopping rule. For any given query scheme  $\fc$, 
querying samples is only stopped when  one of the posteriors becomes larger than $1-\epsilon$, where $\epsilon>0$ is the desired probability of error:
\begin{align}\label{eq:deftau}
\taute:=\min \{t: \max_{i \in \Omega} \rho_i(t)\ge 1-\epsilon\}. 
\end{align}

Let $\mathbb{E}[\tause]$ and $\mathbb{E}[\taute^*]$ denote the optimal expected number of queries in \eqref{Obj} 
and the optimal expected 
number of queries with  the (possibly suboptimal) stopping rule as given in \eqref{eq:deftau}, respectively.  The following fact bounds these quantities both from above and below, and hence 
will be used in Section~\ref{Main} in bounding the loss
of optimality in restricting attention to the above possibly suboptimal stopping rule.  


\begin{lemma}\label{lem:TauvsTildeTau}
Consider stopping times defined earlier with scalars $\iota \geq \epsilon >0$. We have 
\begin{align}\label{tau_vs_tilde_tau}
\mathbb{E} [\tauti] \ (1 - \frac{\epsilon}{\iota}) \leq \mathbb{E}[\tause] \leq \mathbb{E}[ \taute^*].
\end{align}
\end{lemma}

The proof of Lemma~\ref{lem:TauvsTildeTau} is similar to that of Lemma~3 in \cite{EJS-Journal} 
and is given in Appendix~\ref{App:Aux}.

We are now ready to fully describe our proposed heuristic. 
\ignore{ First, recall that restricting our attention to the class of deterministic Markov query strategies does not incur any loss of optimality. As a result, and in the absence of a closed 
form characterization of the optimal policy, we propose the following heuristic (Markov stationary deterministic policy) based on the suboptimal stopping rule given in  \eqref{eq:deftau} and EJS characterization \eqref{EJS-x}: }

\begin{definition}
\label{policy_def}
Policy $\fc_{EJS}$ is a stationary deterministic Markov policy with a suboptimal stopping rule 
defined in (\ref{eq:deftau})
which at a given prior belief $\boldsymbol{\rho}$ queries sample $X_{\fc_{EJS}} \in 
       \argmax \limits_{x \in \mathcal{X}} EJS(\boldsymbol{\rho},x)$.\footnote{Let $\calA$ denote the smallest partition of sample space $\calX$, i.e., $\calX=\cup_{A\in\calA} A$, 
such that for every $A\in\calA$ and $h\in\calH$, the value of $h(x)$ remains constant for all $x\in A$.  
By definition, $EJS(\boldsymbol{\rho},x)=EJS(\boldsymbol{\rho},x')$ for every $x,x'\in A$, $A\in\calA$.
We have $|\calA|\le |\calL|^M$, and hence, $\argmax\limits_{x\in\calX}$ is a valid operation in $\argmax \limits_{x \in \mathcal{X}} EJS(\boldsymbol{\rho},x)$.} \end{definition}


\section{Main Results}
\label{Main}

We now provide the main results -- lower and upper bounds on the optimal number of queries to identify the true function with high accuracy. Note that we expect the query complexity of our problem to depend on the characterizations of the discrete memoryless communication channel (DMC) which corrupts the true label's observations. This is a DMC with input alphabet set 
$\mathcal{L}$, output alphabet set $\mathcal{Y}$, and a collection of conditional probabilities $f_l(\cdot)$, $l\in\mathcal{L}$. We begin with a few assumptions on this channel.

\begin{assumption}
\label{KL0}
$C:=\min \limits_{g\in\mathbb{P}(\calY)} \max \limits_{l \in \mathcal{L}} D(f_{l}\|g)>0$. 
\end{assumption}

\begin{assumption}
\label{c1}
$C_1:= \max \limits_{k, l \in\mathcal{L}} D(f_{k}\|f_{l})< \infty$.
\end{assumption}

\begin{assumption}
\label{Jump}
$C_2 := \max \limits_{k, l \in \mathcal{L}} \sup \limits_{y \in \mathcal{Y}} \frac{f_{k}(y)}{f_{l}(y)} < \infty$.
\end{assumption}

Note that $C$ defined above is nothing but the Shannon capacity of the DMC with the collection of conditional
probabilities  $P(Y=y | L=l) = f_l(y)$, $l\in\mathcal{L}$ (See \cite[Theorem~13.1.1]{CoverBook}). 
In particular, the minimum is achieved by $g^*$, a convex combination of $\{f_l\}_{l\in\mathcal{L}}$, i.e.,
$g^*=\sum_{l\in\calL}\pi^\star_l f_l$ where $\{\pi^\star_l\}_{l\in\mathcal{L}}$
is referred to as the \emph{capacity-achieving input distribution} and has the property that 
for each $k\in\calL$, if $\pi^\star_k > 0$, then $D(f_k \| g^*) = C$ (See \cite[Theorem~4.5.1]{Gallager68}).
If Assumption~\ref{KL0} does not hold, that is if $C=0$, 
the label queries will be completely noisy and no information can be retrieved from the label queries 
regarding the true function. In this sense, Assumption~\ref{KL0} is a necessary condition that ensures 
Problem~(P) has a meaningful solution.

Parameter $C_1$ emerges as an important quantity in the problem of variable-length coding with feedback: 
It denotes the maximum exponential decay rate of the error probability \cite{Burnashev76}. 
It is straight forward to show that $C\le C_1$ and hence, Assumptions~\ref{KL0} and~\ref{c1} imply that also $C_1>0$ and $C<\infty$.

Since, in general, $C_1\le \log C_2$, Assumption~\ref{c1} is redundant with respect to Assumption~\ref{Jump}. 
For observation densities with finite support, i.e., when $|\mathcal{Y}|< \infty$, Assumption~\ref{Jump} ensures that the conditional distributions $f_l$, $l\in\mathcal{L}$, 
are absolutely continuous with respect to each other. Thus for observation densities with finite support, 
Assumption~\ref{Jump} is a necessary and sufficient condition to ensure Assumption~\ref{c1}.  
\ignore{In addition, Assumption~\ref{Jump} is also sufficient to ensure Assumption~\ref{c1} since $C_1\le \log C_2$. In other words,
for observation densities with finite support,  
Assumptions~\ref{c1} and~\ref{Jump} are equivalent. }
On the other hand, for observation kernels with unbounded support, Assumption~\ref{Jump}, which is stronger 
than Assumption~\ref{c1}, is a technical assumption made for notational convenience, and will help 
us construct strong non-asymptotic bounds in closed form.  

While the (non-asymptotic) bounds and analysis in this paper are all obtained under
Assumptions~\ref{KL0} and~\ref{Jump}, we have chosen to separately state Assumptions~\ref{c1} and~\ref{Jump} 
in order to point out that it is possible to relax Assumption~\ref{Jump}. More specifically, it is shown in \cite{HypJournal} that
 at the cost of increasing notation, more complicated analysis, and loosening the non-asymptotic bounds, it is
possible to relax Assumption~\ref{Jump} and obtain similar asymptotic characterizations only under Assumption~\ref{KL0} and
a slightly stronger variant of Assumption~\ref{c1}.

\subsection{Main Results: Lower Bound}

In this subsection,
we show the following lower bound on the minimum expected number of samples required to achieve $\Pe\le\epsilon$.  
\begin{theorem}
\label{LB}
Consider Problem~(P) under Assumptions~\ref{KL0} and~\ref{Jump}. 
\begin{align} \label{LowerBound}
\mathbb{E}[\tause] &\ge 
\left[\frac{(1-\frac{3}{\log\frac{4}{\epsilon}}-\frac{\epsilon}{2}\log\frac{1}{\epsilon})\log M - 2}{C}
+ \frac{\log\frac{1-\epsilon}{\epsilon} - 2\log\log\frac{4}{\epsilon} - \log C_2 - 4}{C_1}\right]^+.
\end{align} 
\end{theorem}

Theorem~\ref{LB} is proved in Appendix~\ref{App:LB} using results in dynamic programming. 
Our lower bound is similar to \cite[Theorem~1]{Burnashev80}, \cite[Theorem~1]{Berlin09}, and \cite[Theorem~6]{Polyanskiy11}.

Next we provide upper bounds on the optimal expected sample size of Bayesian active learning. 

\subsection{Main Results: Upper Bounds}

 In this subsection, we characterize upper bounds on the expected number of sample queries
  in terms of the corresponding Extrinsic Jensen--Shannon (EJS) divergence obtained at each time. In
  our presentation of these results, we will need the following notation:
  \begin{align}
\mathbb{P}^M_\epsilon(\Omega) = \Big\{ \boldsymbol{\rho}\in\mathbb{P}(\Omega): \max \limits_{j\in\Omega} \rho_j\ge \tilde{\rho} \Big\},
  \end{align}
  where 
  \begin{align}
  \label{rhoTildeDef}
  \tilde{\rho} = 1-\frac{1}{1+\max\{\log M,\log\frac{1}{\epsilon}\}}.
  \end{align}  

\begin{theorem}
\label{thm:infoacquisitionUB}
Consider Problem~(P) under Assumptions~\ref{KL0} and~\ref{Jump}.
If there exists a positive value $\alpha$ such that
at any given belief vector $\boldsymbol{\rho}\in \mathbb{P}(\Omega)$, it is possible to find a sample ${x\in\mathcal{X}}$ satisfying 
$EJS(\boldsymbol{\rho},x)\ge\alpha$, then
\begin{align}
\label{UB1stp}
\mathbb{E}[\tause] \le 
\frac{\log M+ \max\{\log\log M,\log\frac{1}{\epsilon}\} +4 C_2}{\alpha}. 
\end{align} 
Furthermore, if there exists a positive value $\beta > \alpha $ such that for all belief vectors 
$\boldsymbol{\rho} \in \mathbb{P}^M_\epsilon(\Omega)$, 
it is possible to find a sample ${x\in\mathcal{X}}$ satisfying 
$EJS(\boldsymbol{\rho},x)\ge\beta$, 
then the following bound is obtained 
\begin{align}
\label{UB2stp}
\mathbb{E}[\tause] \le
\frac{\log M + \max\{\log\log M,\log\log\frac{1}{\epsilon}\}}{\alpha} + \frac{\log\frac{1}{\epsilon}}{\beta} + \frac{3(4C_2)^2}{\alpha\beta}.
\end{align}
\end{theorem}

The proof of the above theorem is constructive and is provided in Appendix~\ref{App:main}. In other words, the policy which selects and queries the label of the sample $x$ 
for which $EJS(\boldsymbol{\rho},x)\ge\alpha$, ensures an expected sample size which is smaller than or equal to the right hand side of (\ref{UB1stp}). Now, by construction, policy $\fc_{EJS}$ is such a policy.  A similar statement holds for (\ref{UB2stp}). 

We remark that as $\beta$ is the minimum value of $EJS(\boldsymbol{\rho}, x)$ over a subset of belief vectors $\boldsymbol{\rho} \in \mathbb{P}_\epsilon^M(\Omega)$, and $\alpha$ is the minimum value over all belief vectors, $\beta \geq \alpha$, \eqref{UB2stp} illustrates that we can get significantly better bounds when $\beta$ is much greater than $\alpha$.

\subsection{Main Results: Asymptotic and Order Optimality}

Note that the lower and upper bounds provided by Theorems~\ref{LB} and~\ref{thm:infoacquisitionUB}
are non-asymptotic and hold for all values of $M$ and $\epsilon$. Nonetheless, they 
can be applied to establish the asymptotic and order optimality 
of $\fc_{EJS}$ as defined in Section \ref{Asym}:

\begin{corollary}
The proposed Markov deterministic heuristic policy which maximizes the EJS divergence
 is order optimal in $\epsilon$ and $M$ if there exists scalar $\alpha>0$ satisfying the first condition of Theorem~\ref{thm:infoacquisitionUB} such that $\alpha \not\to0$ as $M\to\infty$ or $\epsilon\to 0$.
 Furthermore, it is asymptotically optimal in $\epsilon$ (and $M$) if $\beta$ can be selected to be as large as $C_1$ (and $\alpha$ as large as $C$). 
\end{corollary}


However, the above results depend on characterizing non-zero values, if not sufficiently large values, for quantities $\alpha$ and $\beta$, 
which in turn depend on the function class $\mathcal{H}$ and the set of samples that we are allowed to pick from. 
In the next subsection, we specialize the above results to several function classes in order to concretely illustrate the 
asymptotic performance of $\fc_{EJS}$.

\subsection{Applications and Consequences} 

So far, we have only characterized the performance of $\fc_{EJS}$ in terms of strictly positive scalars $\alpha$ and $\beta$, assuming they do exist. An important question
remains as whether one can always find such scalars. 
In this section, we specifically look at an important function class example
 and provide nontrivial characterization of $\alpha$ and $\beta$, hence, 
demonstrating the relative looseness/tightness of the upper bounds.  Furthermore, we discuss the asymptotic and order optimality of these bounds.

We begin with the following definitions which will 
allow us to generalize the notion of \mbox{1-neighborly}, first suggested by \cite{Nowak09}; then
for this general class,  we will obtain non-trivial scalars $\alpha$ and $\beta$ satisfying the 
conditions of Theorem~\ref{thm:infoacquisitionUB}. 

Consider the representation of a pair of samples $x$ and $x'$ in terms of their partitioning of the functions:

\begin{definition}
A pair of samples $x, x' \in \mathcal{X}$ partition the function class $\mathcal{H}$ in an agreement set
 $A_{x,x'}:=\{h \in \mathcal{H}:  h(x)=h(x')\}$ and a disagreement set  $\Delta_{x,x'}:=\{h \in \mathcal{H}:  h(x)\neq h(x')\}$.
\end{definition}

\begin{definition}
A class of functions $\mathcal{H}$ is referred to as locally identifiable
if for any $h_i\in\mathcal{H}$, there exist samples $x,x'\in\mathcal{X}$ 
and labels $l,l'\in\mathcal{L}$ such that either of the following be true
\ignore{\begin{align*}
[h_i(x),h_i(x')]=[l,l'], \quad \text{and} \quad [h_j(x),h_j(x')] \in \bigcup_{k\in\mathcal{L}}\{[k,k]\} \cup \{[l',l]\}, \quad \forall j \neq i;
\end{align*}
or
\begin{align*}
[h_i(x),h_i(x')]=[l,l], \quad \text{and} \quad [h_j(x),h_j(x')] \in \{[l,l'],[l',l],[l',l']\}, \quad \forall j \neq i.
\end{align*}}
\begin{enumerate}
  \item[$(i)$]  $h_i \in \Delta_{x,x'} \cap H^x_l \cap H^{x'}_{l'}$ and $\mathcal{H}-\{h_i\} =  A_{x,x'} \cup (H^x_{l'} \cap H^{x'}_{l})$, or
  \item [$(ii)$] $\{h_i\} = A_{x,x'} \cap H^x_l$ and for all $k \neq l, l'$, $H^x_k \cup H^{x'}_k = \emptyset$. 
\end{enumerate}
\end{definition}

In essence, the locally identifiable condition implies that for any function $h_i\in\mathcal{H}$, 
there are (at least) two samples $x$ and $x'$ in~$\mathcal{X}$ and two labels $l$ and $l'$ using which $h_i$ can be distinguished from all other functions. As we will see in Section~\ref{SC}, local identifiability is a fairly mild condition that is satisfied by a number of natural function classes. 	


The performance of $\fc_{EJS}$ when the labeling function class is locally identifiable is characterized by
 the capacity of the (sub)channel with two inputs $l,l' \in\mathcal{L}$ denoted by $C_{ll'}$, i.e.,
\begin{align}
\label{capacity2}
C_{ll'} &:= \min \limits_{g\in\mathbb{P}(\calY)} \max \{D(f_l\|g),D(f_{l'}\|g)\}, 
\end{align}
and consequently 
\begin{align}
\label{cmin}
C_\tmin := \min \limits_{l,l'\in\mathcal{L}, l\neq l'} \min \left\{ C_{ll'} , D\left(f_{l'} \left\| \frac{1}{2} f_l + \frac{1}{2} f_{l'} \right. \right) \right\}.
\end{align}		

\begin{proposition}
\label{gen1nbr}
When function class $\mathcal{H}$ is locally identifiable, $\alpha\geq\frac{1}{M} C_\tmin$ and $\beta\geq\tilde{\rho} C_\tmin$. 
More precisely,  for every belief vector $\boldsymbol{\rho}$, there
exists an $x\in \mathcal{X}$ such that \begin{align}
EJS(\boldsymbol{\rho},x) \geq \begin{cases}  \frac{1}{M} C_\tmin &\mbox{if } \boldsymbol{\rho} \notin \mathbb{P}^M_\epsilon(\Omega) \\
\tilde{\rho} C_\tmin &\mbox{otherwise}
 \end{cases}. 
\end{align}
\end{proposition}


\begin{IEEEproof}
To prove Proposition~\ref{gen1nbr}, it suffices to show that
$$\max_{x\in\mathcal{X}} EJS(\boldsymbol{\rho},x) \ge \max_{i\in\Omega} \rho_i C_\tmin.$$
	
Let $\hat{i}=\argmax \limits_{i\in\Omega} \rho_i$. 
By definition of the locally identifiable class, there exist $x_{\hat{i}},x'_{\hat{i}}\in\mathcal{X}$ and $l,l'\in\mathcal{L}$ such that one of the following conditions holds
\begin{align}
\label{LocId1}
[h_{\hat{i}}(x_{\hat{i}}),h_{\hat{i}}(x'_{\hat{i}})]=[l,l'] \ \ \text{and} \ \ [h_j(x_{\hat{i}}),h_j(x'_{\hat{i}})] \in \bigcup_{k\in\mathcal{L}}\{[k,k]\} \cup \{[l',l]\}, \ \ \forall j \neq \hat{i},\\
\label{LocId2}
[h_{\hat{i}}(x_{\hat{i}}),h_{\hat{i}}(x'_{\hat{i}})]=[l,l] \quad \text{and} \quad [h_j(x_{\hat{i}}),h_j(x'_{\hat{i}})] \in \{[l,l'],[l',l],[l',l']\}, \quad \forall j \neq \hat{i}.
\end{align}
 
For any $k,k'\in\mathcal{L}$, let 
$$\pi_{kk'}:=\sum_{j\in\Omega \colon [h_j(x_{\hat{i}}),h_j(x'_{\hat{i}})]=[k,k']} \frac{\rho_j}{1-\rho_{\hat{i}}}.$$
	
	Suppose \eqref{LocId1} holds. Then
	\begin{align}	
	\label{LocId1EJS}
	\nonumber
	\lefteqn{\max_{x\in\mathcal{X}} EJS(\boldsymbol{\rho},x)}\\ \nonumber 
	&\ge \max \left\{EJS(\boldsymbol{\rho},x_{\hat{i}}), EJS(\boldsymbol{\rho},x'_{\hat{i}})\right\}\\ \nonumber
	&\ge \rho_{\hat{i}} \max \Bigg\{\hspace*{-.025in}D\bigg(f_{h_{\hat{i}}(x_{\hat{i}})} \| \sum_{j\neq\hat{i}} \frac{\rho_j}{1-\rho_{\hat{i}}} f_{h_j(x_{\hat{i}})} \bigg) , D\bigg(f_{h_{\hat{i}}(x'_{\hat{i}})} \| \sum_{j\neq\hat{i}} \frac{\rho_j}{1-\rho_{\hat{i}}}f_{h_j(x'_{\hat{i}})} \bigg) \hspace*{-.025in} \Bigg\}\\ \nonumber
	&= \rho_{\hat{i}} \max \Bigg\{ \hspace*{-.025in}D \bigg(f_l \| \sum_{k\in\mathcal{L}} \pi_{kk} f_k + \pi_{l'l} f_{l'}\bigg),
	D\bigg(f_{l'} \| \sum_{k\in\mathcal{L}} \pi_{kk} f_k + \pi_{l'l} f_{l}\bigg) \hspace*{-.025in}\Bigg\}\\ \nonumber
	&\stackrel{(a)}{\ge} \rho_{\hat{i}} \max \Bigg\{ \hspace*{-.025in} D \bigg(f_l \| \frac{\sum\limits_{k\in\mathcal{L}} \pi_{kk} f_k + \pi_{l'l} f_{l'} + \pi_{l'l} f_{l}}{1+\pi_{l'l}}\bigg),
	D\bigg(f_{l'} \| \frac{\sum\limits_{k\in\mathcal{L}} \pi_{kk} f_k + \pi_{l'l} f_{l} + \pi_{l'l} f_{l'}}{1+\pi_{l'l}}\bigg) \hspace*{-.025in} \Bigg\} \\ \nonumber
	&\ge \rho_{\hat{i}} \min_g \max\{D(f_l\|g),D(f_{l'}\|g) \}\\ \nonumber
	&= \rho_{\hat{i}} C_{ll'}\\ 
	&\ge \max_{i\in\Omega} \rho_i C_\tmin,
	\end{align}
	where $(a)$ follows by Fact~\ref{DPQa} in Appendix~\ref{App:Aux}. 
	
On the other hand, if \eqref{LocId2} holds, then
	\begin{align}
	\label{LocId2EJS}	
	\nonumber
	\lefteqn{\max_{x\in\mathcal{X}} EJS(\boldsymbol{\rho},x)}\\ \nonumber 
	&\ge \rho_{\hat{i}} \max \Big\{ D \big(f_l \| \pi_{ll'} f_l + (\pi_{l'l}+\pi_{l'l'}) f_{l'}\big),
	D\big(f_{l} \| \pi_{l'l} f_l + (\pi_{ll'}+\pi_{l'l'}) f_{l'} \big) \Big\}\\ \nonumber
	&\stackrel{(a)}{\ge} \rho_{\hat{i}} D \Big(f_l \|  \frac{1}{2} f_l + \frac{1}{2} f_{l'}\Big)\\ 
	&\ge \max_{i\in\Omega} \rho_i C_\tmin,
	\end{align}
	where $(a)$ follows by Fact~\ref{DPQa} in Appendix~\ref{App:Aux} and since $\min\{\pi_{ll'},\pi_{l'l}\}\le\frac{1}{2}$.

Combining \eqref{LocId1EJS} and \eqref{LocId2EJS}, we have the assertion of the proposition.
\end{IEEEproof}

The following corollary provides an upper bound on the 
expected number of sample queries. 

\begin{corollary}
Consider Problem~(P) under Assumptions~\ref{KL0} and~\ref{Jump}. If the function class $\mathcal{H}$ is locally identifiable, then
\begin{align} \label{LocIdUB}
\mathbb{E}[\tause] \le 
\frac{M(\log M + \max\{\log\log M,\log\log\frac{1}{\epsilon}\})}{C_{\min}} + \frac{\log\frac{1}{\epsilon}}{\tilde{\rho} C_{\min}} + \frac{3M(4C_2)^2}{\tilde{\rho} C_{\min}^2}.
\end{align}
\end{corollary}

Next, we define a subclass of the locally identifiable function class, and show that for this function class,
 $\alpha$ and $\beta$ can be selected to match the denominators in the lower bound in \eqref{LowerBound}. Hence,
the policy $\fc_{EJS}$ is  provably asymptotically optimal in $\epsilon$ and~$M$.

\begin{definition}
We call the function class $\mathcal{H}$ $\mathcal{R}(\mathcal{H})$-sample-rich for $\mathcal{R}(\mathcal{H}) = \cup_{x\in\calX} \Xi^x$. 
In the special case where
$\mathcal{R}(\mathcal{H})$ includes all ($|\mathcal{L}|^M -|\mathcal{L}|$) non-trivial $|\calL|$-partitions of $\mathcal{H}$, we simply refer to $\mathcal{H}$ as sample-rich. \end{definition}

\begin{proposition}\label{lem:samplerich}
When function class $\mathcal{H}$ is sample-rich, $\alpha\geq C$ and $\beta\geq\tilde{\rho} C_1$. 
\end{proposition}
\begin{IEEEproof}
To prove Proposition~\ref{lem:samplerich}, we will show that for all belief vectors $\boldsymbol{\rho}$,
$$
\max_{x\in\mathcal{X}}EJS(\boldsymbol{\rho},x) \geq C,
$$
and furthermore,
$$
\max_{x\in\mathcal{X}}EJS(\boldsymbol{\rho},x) \geq \max_{i\in\Omega} \rho_i C_1.
$$

Recall from Section~\ref{Main} that 
\begin{align}
\label{DefCg} 
C=\min \limits_{g\in\mathbb{P}(\calY)} \max \limits_{l \in \mathcal{L}} D(f_{l}\|g),
\end{align}
and the minimum is achieved by $g^*=\sum_{l\in\calL}\pi^\star_l f_l$ where $\pi^\star$ is the capacity achieving 
input distribution, i.e., 
\begin{align}
\label{LGal}
D\bigg(f_k \Big\| \sum \limits_{l\in\mathcal{L}} \pi^\star_l f_l\bigg) = C \quad \text{for any \ $k\in\mathcal{L}$ \ such that \ $\pi^\star_k > 0$}.
\end{align}

By definition of the sample-rich function class, for each $\boldsymbol{v}:= [v_1, \ldots, v_M] \in\mathcal{L}^M$, there exists a sample in $\mathcal{X}$, say~$x_{\boldsymbol{v}}$, that satisfies $\boldsymbol{h}(x_{\boldsymbol{v}})=\boldsymbol{v}$, 
where $\boldsymbol{h}(x) := [h_1(x),h_2(x),\ldots,h_{M}(x)]$.
Let
\begin{align*}
\lambda^{\star}_{\boldsymbol{v}}= \prod_{i=1}^M \pi^\star_{v_i}. 
\end{align*}
Note that $\sum_{\boldsymbol{v}\in\mathcal{L}^M} \lambda^{\star}_{\boldsymbol{v}}=1$. 
Moreover, 
for any~$i,j \in \Omega$, $i \neq j$,
$$\sum_{\boldsymbol{v}\in\mathcal{L}^M \colon v_i=k} \lambda^{\star}_{\boldsymbol{v}} = \pi^\star_k, \ \ \ 
\sum_{\boldsymbol{v}\in\mathcal{L}^M \colon v_i=k, v_j=l} \lambda^{\star}_{\boldsymbol{v}} = \pi^\star_k \pi^\star_l.$$

Using weights $\{\lambda^{\star}_{\boldsymbol{v}}\}_{\boldsymbol{v}\in\mathcal{L}^M}$ and taking average over all $\boldsymbol{v}\in\mathcal{L}^M$, we obtain
	\begin{align*}	
\max_{x\in\mathcal{X}} EJS(\boldsymbol{\rho},x)
	&\ge \sum_{\boldsymbol{v}} \lambda^{\star}_{\boldsymbol{v}} EJS(\boldsymbol{\rho},x_{\boldsymbol{v}})\\
	&= \sum_{\boldsymbol{v}} \lambda^{\star}_{\boldsymbol{v}} \sum_{i=1}^M \rho_i D\bigg(f_{h_i(x_{\boldsymbol{v}})}\big\|\sum_{j\neq i} \frac{\rho_j}{1-\rho_i} f_{h_j(x_{\boldsymbol{v}})}\bigg)\\
	&= \sum_{i=1}^M \rho_i  \sum_{k\in\mathcal{L}} \pi^\star_k \sum_{\boldsymbol{v} \colon v_i=k} \frac{\lambda^{\star}_{\boldsymbol{v}}}{\pi^\star_k} D\bigg(f_k \big\| \sum_{j\neq i} \frac{\rho_j}{1-\rho_i} f_{v_j} \bigg) \\
	&\stackrel{(a)}{\ge} \sum_{i=1}^M \rho_i  \sum_{k\in\mathcal{L}} \pi^\star_k  
	D\bigg(f_k \big\| \sum_{j\neq i} \frac{\rho_j}{1-\rho_i} \sum_{\boldsymbol{v} \colon v_i=k} \frac{\lambda^{\star}_{\boldsymbol{v}}}{\pi^\star_k} f_{v_j} \bigg) \\
	&= \sum_{i=1}^M \rho_i  \sum_{k\in\mathcal{L}} \pi^\star_k  
	D\bigg(f_k \big\| \sum_{j\neq i} \frac{\rho_j}{1-\rho_i} \sum_{l\in\mathcal{L}} \sum_{\boldsymbol{v} \colon v_i=k, v_j=l} \frac{\lambda^{\star}_{\boldsymbol{v}}}{\pi^\star_k} f_l \bigg) \\
	&= \sum_{i=1}^M \rho_i  \sum_{k\in\mathcal{L}} \pi^\star_k  
	D\bigg(f_k \big\| \sum_{l\in\mathcal{L}} \pi^\star_l f_l \bigg) \\
	&\stackrel{(b)}{=} \sum_{i=1}^M \rho_i C\\
	&= C,
	\end{align*}
	where $(a)$ follows from Jensen's inequality and $(b)$ follows from \eqref{LGal}.

Let $\hat{i}=\argmax \limits_{i\in\Omega} \rho_i$.
Let $k,l \in \mathcal{L}$ be the labels satisfying $D(f_k\|f_l)=C_1$. 
By definition of the sample-rich function class, there exists a sample $x_{\hat{i}} \in\mathcal{X}$ that satisfies $h_{\hat{i}}(x_{\hat{i}})=k$ and $h_j(x_{\hat{i}})=l$ for all $j\neq i$. We have
	\begin{align*}	
\max_{x\in\mathcal{X}} EJS(\boldsymbol{\rho},x) \ge
	EJS(\boldsymbol{\rho},x_{\hat{i}})
	\ge \rho_{\hat{i}} D\bigg(f_{h_{\hat{i}}(x_{\hat{i}})} \Big\| \sum_{j\neq\hat{i}} \frac{\rho_j}{1-\rho_{\hat{i}}} f_{h_j(x_{\hat{i}})} \bigg) 
	= \max_{i\in\Omega} \rho_i C_1.
	\end{align*}
\end{IEEEproof}

As a simple corollary, 
\begin{corollary}
Consider Problem~(P) under Assumptions~\ref{KL0} and~\ref{Jump}. If the function class $\mathcal{H}$ is
sample-rich, 
\begin{align}
\mathbb{E}[\tause] \le \frac{\log M + \max\{\log \log M, \log \log \frac{1}{\epsilon}\}}{C} + \frac{\log \frac{1}{\epsilon}}{\tilde{\rho} C_1} + \frac{48 C_2^2}{\tilde{\rho} C C_1}.
\end{align} 
\end{corollary}

The above results show that for sample-rich function classes, $\fc_{EJS}$ is asymptotically optimal in both $\epsilon$ and $M$. 


The above results generalize the finding of \cite{Nowak09} to a multi-label Bayesian learning with non-binary and asymmetric noise case. However, to make this comparison precise, we will dedicate the next section to specialize our general results above to the noisy generalized binary search of \cite{Nowak09}.

\section{Special Case: Noisy Generalized Binary Search}
\label{SC}


We next compare our work with existing results. Since the only study of similar nature is that of noisy generalized binary search \cite{Nowak09}, we  consider an application of our main results to noisy generalized binary search among 1-neighborly functions, first introduced in \cite{Nowak09}. This is a special case of our problem where functions are binary-valued, i.e., $\mathcal{L}=\{-1,+1\}$, the observation space $\calY = \{-1, +1\}$, and observation densities are of the following form: 
 \[f_l(y)=\left\{\begin{array}{ll} 1-p & \mbox{if } y=l \\
					       p & \mbox{if } y=-l \end{array} \right. ,\]
for some $p\in (0,1/2)$. In other words, for any sample $x$, if $h_i$ is the true function, then the label $h_i(x)$ is observed through a binary symmetric channel with crossover probability $p$.

For the case of noisy generalized binary search, $C$, $C_1$, and $C_2$ defined in Section~\ref{Main}
can be further simplified to
\begin{align*}
C &:= 1+p\log p+(1-p)\log(1-p),\\
C_1 &:= p\log\frac{p}{1-p}+(1-p)\log\frac{1-p}{p},\\
C_2 &:= \frac{1-p}{p}.
\end{align*}
In order to emphasize the dependence of $C$, $C_1$, and $C_2$ on the Bernoulli parameter $p$ (corresponding to the observation noise),
we denote them by $C(p)$, $C_1(p)$, and $C_2(p)$ respectively.
Note that from Jensen's inequality, $C_1(p)\ge2C(p)$.


Next we define a class of 1-neighborly functions first defined in \cite[Definition~2]{Nowak09}. 

\begin{definition}
A class of binary-valued functions $\mathcal{H}$ is referred to as 1-neighborly
if for any $h_i\in\mathcal{H}$, there exist $x,x'\in\mathcal{X}$ such that
\begin{align*}
	 \left\{ \begin{array}{ll}
       h_i(x) \neq h_i(x') & \\
       h_j(x) = h_j(x') & \ \ \text{if $j \neq i$ and $h_j(\cdot) \neq -h_i(\cdot)$}                    \end{array} \right. .
  \end{align*}
\end{definition}

It is simple to see that the class of 1-neighborly functions is a subset of binary-valued locally identifiable function class. This 
implies the following baseline bound:

\begin{corollary}
\label{bin1nbr}
When function class $\mathcal{H}$ is 1-neighborly, 
we have $\alpha \geq \frac{1}{M} C(p)$ and $\beta \geq \tilde{\rho} C(p)$.\end{corollary}

In comparison, \cite{Nowak09} provides two sample query strategies, NGBS and MSGBS, 
whose performance (upper bound) depends strongly on the properties of the function class at hand. 

Let $n_0$ denote the number of queries made by GBS to determine $h_{\theta}$ in the noiseless setting.
The number of queries required by NGBS to attain $\Pe\le\epsilon$ is upper bounded by 
\begin{align}
\label{NowakUB2}
\frac{n_0(\log n_0 + \log\frac{1}{\epsilon})}{(\frac{1}{2}-p)^2}.
\end{align}

Let $\calA$ denote the smallest partition of sample space $\calX$, i.e., $\calX=\cup_{A\in\calA} A$, 
such that for every $A\in\calA$ and $h\in\calH$, the value of $h(x)$ is constant for all $x\in A$;
and denote this value by $h(A)$.
Furthermore, let 
\begin{align}
\label{c*def}
c^* &:= \min\limits_{P\in\mathbb{P}(\calA)} \max\limits_{h\in\calH} \left|\sum_{A\in\calA} h(A) P(A)\right|.
\end{align}

Under MSGBS, the number of queries required to ensure that $\Pe\le\epsilon$ 
is upper bounded by 
\begin{align}
\label{NowakUB}
\frac{\log M + \log\frac{1}{\epsilon}}{\min\{2(1-c^*),1\} \lambda(p)},
\end{align}
where
\begin{align}
\label{lambdap}
\lambda(p) &:= \max\limits_{p'\in (p,1/2)} \frac{1}{4}\left(1-\frac{p'(1-p)}{1-p'}-\frac{(1-p')p}{p'}\right). 
\end{align}


Note that $c^*$ (as well as $n_0$) in general depends on the function class $\mathcal{H}$. 
Since this dependence is implicit and hard to 
characterize in closed form for general function class $\mathcal{H}$, a direct comparison between (\ref{NowakUB})
(or (\ref{NowakUB2})) and (\ref{LocIdUB})  is not possible. 
As a result, next we focus on special cases of function classes studied in \cite{Nowak09} for which a precise characterization of 
the achievable upper bound is available. 
Consequently, we next define two important subclasses of 1-neighborly binary-valued functions:
1) Disjoint class $\mathcal{H}_{D}$;
2) Threshold class $\mathcal{H}_{T}$. 
We further specialize the choices of $\alpha$ and $\beta$ for these classes.

\begin{definition}
Let ${\boldsymbol{e}}_{i}$, $i\in\Omega$, represent a vector of size $M$ whose $i^{\text{th}}$ element is $+1$ and all other elements are $-1$.
A collection of functions $\mathcal{H}$ is referred to as \emph{disjoint interval class} if
$\cup_{x\in\mathcal{X}} \{\boldsymbol{h}(x)\}=\cup_{i\in\Omega} \{\boldsymbol{e}_i\} \subset \{-1,+1\}^M$,
where $\boldsymbol{h}(x) := [h_1(x),h_2(x),\ldots,h_{M}(x)]$.
In other words, for any sample $x\in\mathcal{X}$, only one function in $\mathcal{H}$ takes value $+1$ and all other functions take value $-1$.
\end{definition}

\begin{definition}
Let ${\boldsymbol{u}}_{i}$, $i\in\Omega$, represent a vector of size $M$ whose first $i$ elements are $-1$ and all other elements are $+1$.
A collection of functions $\mathcal{H}$ is referred to as \emph{threshold class} if
$\cup_{x\in\mathcal{X}} \{\boldsymbol{h}(x)\}=\cup_{i\in\Omega} \{\boldsymbol{u}_i\} \subset \{-1,+1\}^M$.
\end{definition}

\begin{fact}[see \cite{Nowak11IT}]
\label{fact:NowakParams}
For the disjoint interval class $\calH_D$, $n_0 \le M$ and $c^*=1-\frac{2}{M}$.
For the threshold function class $\calH_T$, $n_0 \le \log M$ and $c^*=0$.
For the sample-rich function class $\calH_R$, $n_0 \le \log M$ and $c^*=0$.
\end{fact} 
 
 We are now ready to contrast these results with our findings. In particular, we have 


\begin{proposition}
\label{prop:binary}
For the disjoint interval class $\calH_D$, $\alpha\geq\frac{1}{M} C_1(p)$ and $\beta\geq\tilde{\rho} C_1(p)$.
For the threshold function class $\calH_T$, $\alpha \geq C(p)$ and $\beta \geq C(p)$.
For the sample-rich function class $\calH_R$, $\alpha \geq C(p)$ and $\beta \geq \tilde{\rho} C_1(p)$.
\end{proposition}

The proof of Proposition~\ref{prop:binary} is provided in Appendix~\ref{app:propbinary}.


\ignore{\subsection{Comparison to Known Results}

In this section, we compare the performance of $\fc_{EJS}$
to that of NGBS and MSGBS policies proposed in \cite{Nowak11IT}. 

\subsubsection{Theoretical Comparison to Existing Results}}
Table~\ref{SC_T2} summarizes our results and specializes the upper bounds in \cite{Nowak11IT} and lists 
the number of samples required by the policies NGBS, MSGBS, and $\fc_{EJS}$ to attain $\Pe\le\epsilon$.
Furthermore, these bounds 
together with \eqref{asymLB} establish asymptotic and order optimality of $\fc_{EJS}$.\footnote{The term $o(1)$ goes to zero as $\epsilon\to 0$ or $M\to\infty$. See Appendix~\ref{o1to0} for more details.} 

Recall that policies NGBS and MSGBS are non-sequential in the sense that they stop after a fixed number of samples, regardless of the probability of error. The numbers shown in Table~\ref{SC_T2} are the number of samples that these policies require to achieve $\Pe\le\epsilon$. Policy $\fc_{EJS}$ is sequential and Table~\ref{SC_T2} shows the expected number of samples required by this policy to achieve $\Pe\le\epsilon$.

\begin{table}[htp]
\center
\caption{Performance comparison of NGBS, MSGBS, and $\fc_{EJS}$ on different function classes.}
\label{SC_T2}
\begin{tabular}{cccc}
  \toprule
  Function class & NGBS & MSGBS & $\fc_{EJS}$ \\
  \midrule
  \vspace{0.05 in} 
  Disjoint $\mathcal{H}_{D}$ & $\frac{M (\log M + \log\frac{1}{\epsilon})}{(\frac{1}{2}-p)^2}$ & $\frac{M(\log M+\log\frac{1}{\epsilon})}{4 \lambda(p)}$ & $\left(\frac{M \log M}{C_1(p)} + \frac{\log\frac{1}{\epsilon}}{C_1(p)}\right)(1+o(1))$\\
  & order optimal in $\epsilon$ & order optimal in $\epsilon$ & asymptotic optimal in $\epsilon$\\
  \midrule
  \vspace{0.05 in} 
  Threshold $\mathcal{H}_{T}$ & $\frac{\log M(\log\log M + \log\frac{1}{\epsilon})}{(\frac{1}{2}-p)^2}$ & $\frac{\log M+\log\frac{1}{\epsilon}}{\lambda(p)}$ & $\left(\frac{\log M}{C(p)} + \frac{\log\frac{1}{\epsilon}}{C(p)}\right)(1+o(1))$\\
  & order optimal in $\epsilon$ & order optimal in $\epsilon, M$ & order optimal in $\epsilon, M$\\
  \midrule
  \vspace{0.05 in}
  Sample-rich $\mathcal{H}_{R}$ & $\frac{\log M(\log\log M + \log\frac{1}{\epsilon})}{(\frac{1}{2}-p)^2}$ & $\frac{\log M+\log\frac{1}{\epsilon}}{\lambda(p)}$ & $\left(\frac{\log M}{C(p)} + \frac{\log\frac{1}{\epsilon}}{C_1(p)}\right)(1+o(1))$\\
 & order optimal in $\epsilon$ & order optimal in $\epsilon, M$ & asymptotic optimal in $\epsilon, M$\\
  \bottomrule
\end{tabular}
\end{table}


To provide a comparison between the obtained bounds, in asymptotic regime, Fig.~\ref{fig:CC1b} compares the denominators of the upper bounds given in Table~\ref{SC_T2}. 
Note that our upper bound provides improvement over those corresponding to NGBS and MSGBS. 
Particularly, the gap between the bounds is very significant for small values of the Bernoulli parameter $p$
and for large values of $\frac{1}{\epsilon}$ and $M$. 

%
%

\begin{figure}[htp]
\centering 
\psfrag{P}{\scriptsize{$p$}}
\psfrag{C1}{\scriptsize{$C_1(p)$}}
\psfrag{CC}{\scriptsize{$C(p)$}}
\psfrag{NGGG}{\scriptsize{$(\frac{1}{2}-p)^2$}}
\psfrag{LA}{\scriptsize{$\lambda(p)$}}

\includegraphics[width=0.875\textwidth]{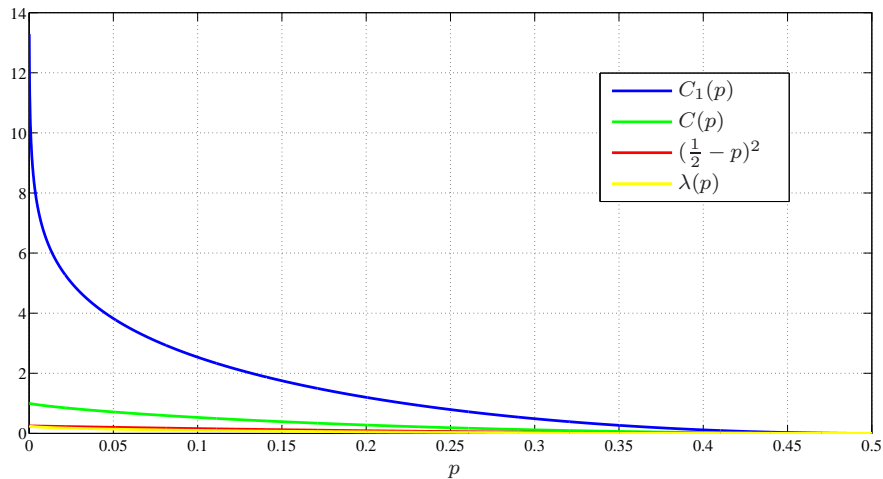}
\caption{Comparison of $C(p)$, $C_1(p)$, $(\frac{1}{2}-p)^2$, and $\lambda(p)$, for $p\in(0,1/2)$.}
\label{fig:CC1b}
\end{figure}

\begin{remarks} 
With no tight lower bound on the performance of NGBS and MSGBS, the above comparison 
must not be confused with a comparative analysis between $\fc_{EJS}$  versus NGBS and MSGBS. In fact, 
the gap between the above upper bounds  could potentially be due to the analysis limitation in \cite{Nowak11IT} of these algorithms rather than their performance.  
\end{remarks}

Next, policies $\fc_{EJS}$ and MSGBS are compared numerically for the problem
of noisy generalized binary search with parameter $p$ and a rich function class of size $M$ (we do not consider NGBS since it is outperformed by MSGBS). This numerical study not only sheds light on non-asymptotic performance of 
both policies but also provides a direct comparison between the performance of these policies (as opposed to a comparison 
between the upper bounds on the performance of these policies given in Table~I). 

In order to have a fair comparison, the candidate policies are compared in both sequential and non-sequential scenarios. 
In the sequential scenario, the policies stop as soon as the belief about one of the functions passes a threshold $1-\epsilon$,
and the expected number of queries is considered as a measure of performance;
while in the non-sequential scenario, the policies are compared based on their average probability of making a wrong declaration 
after $N$ number of label queries.

Figs.~\ref{fig:eps} and~\ref{fig:M} show the performance of $\fc_{EJS}$ and MSGBS for the sequential scenario 
while Figs.~\ref{fig:nonN} and~\ref{fig:nonM} compare their performance for the non-sequential scenario. 
%

\begin{figure}[htp]
\centering 
\psfrag{eps}{\scriptsize{$\epsilon$}}
\psfrag{Num}{\hspace*{.15in} \scriptsize{Expected number of samples}}
\psfrag{MSGBS}{\scriptsize{MSGBS}}
\psfrag{EJS}{\scriptsize{$\fc_{EJS}$}}
\includegraphics[width=0.875\textwidth]{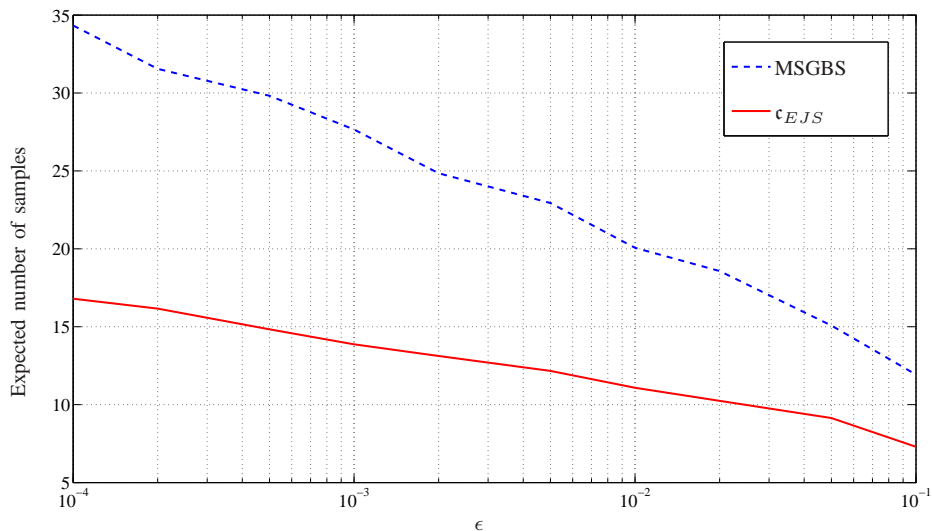}
\caption{Sequential noisy generalized binary search with parameter $p=0.2$, desired probability of error $\epsilon$, and a rich function class of size $M=5$. The expected number of samples is plotted as $\epsilon$ varies.}
\label{fig:eps}
\end{figure}

\begin{figure}[htp]
\centering 
\psfrag{M}{\scriptsize{$M$}}
\psfrag{Num}{\hspace*{.15in} \scriptsize{Expected number of samples}}
\psfrag{MSGBS}{\scriptsize{MSGBS}}
\psfrag{EJS}{\scriptsize{$\fc_{EJS}$}}
\includegraphics[width=0.875\textwidth]{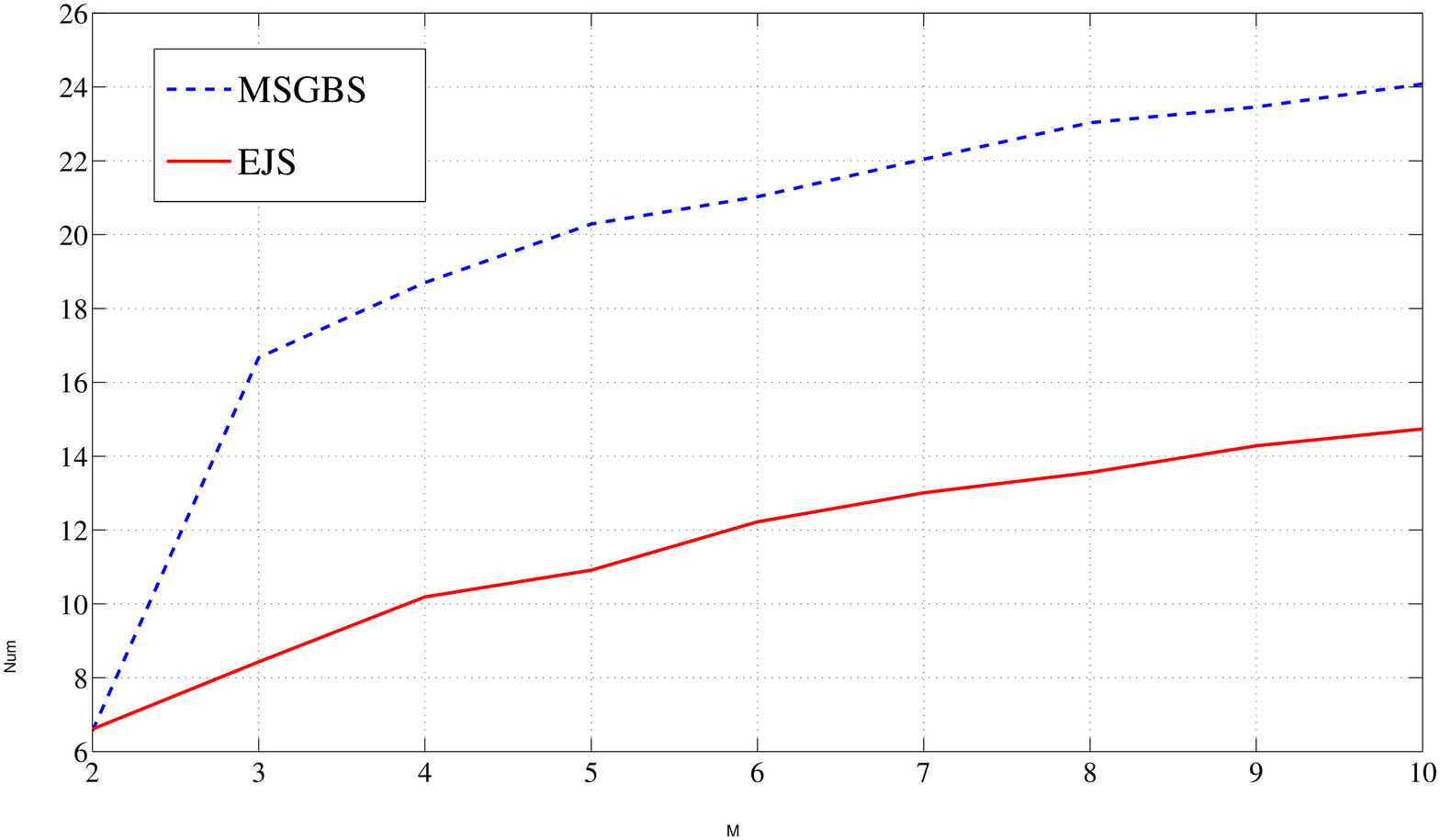}
\caption{Sequential noisy generalized binary search with parameter $p=0.2$, desired probability of error $\epsilon=0.01$, and a rich function class of size $M$. The expected number of samples is plotted as $M$ varies.}
\label{fig:M}
\end{figure}

\begin{figure}[htp]
\centering 
\psfrag{N}{\scriptsize{$N$}}
\psfrag{Pe}{\scriptsize{Pe}}
\psfrag{MSGBS}{\scriptsize{MSGBS}}
\psfrag{EJS}{\scriptsize{$\fc_{EJS}$}}
\includegraphics[width=0.875\textwidth]{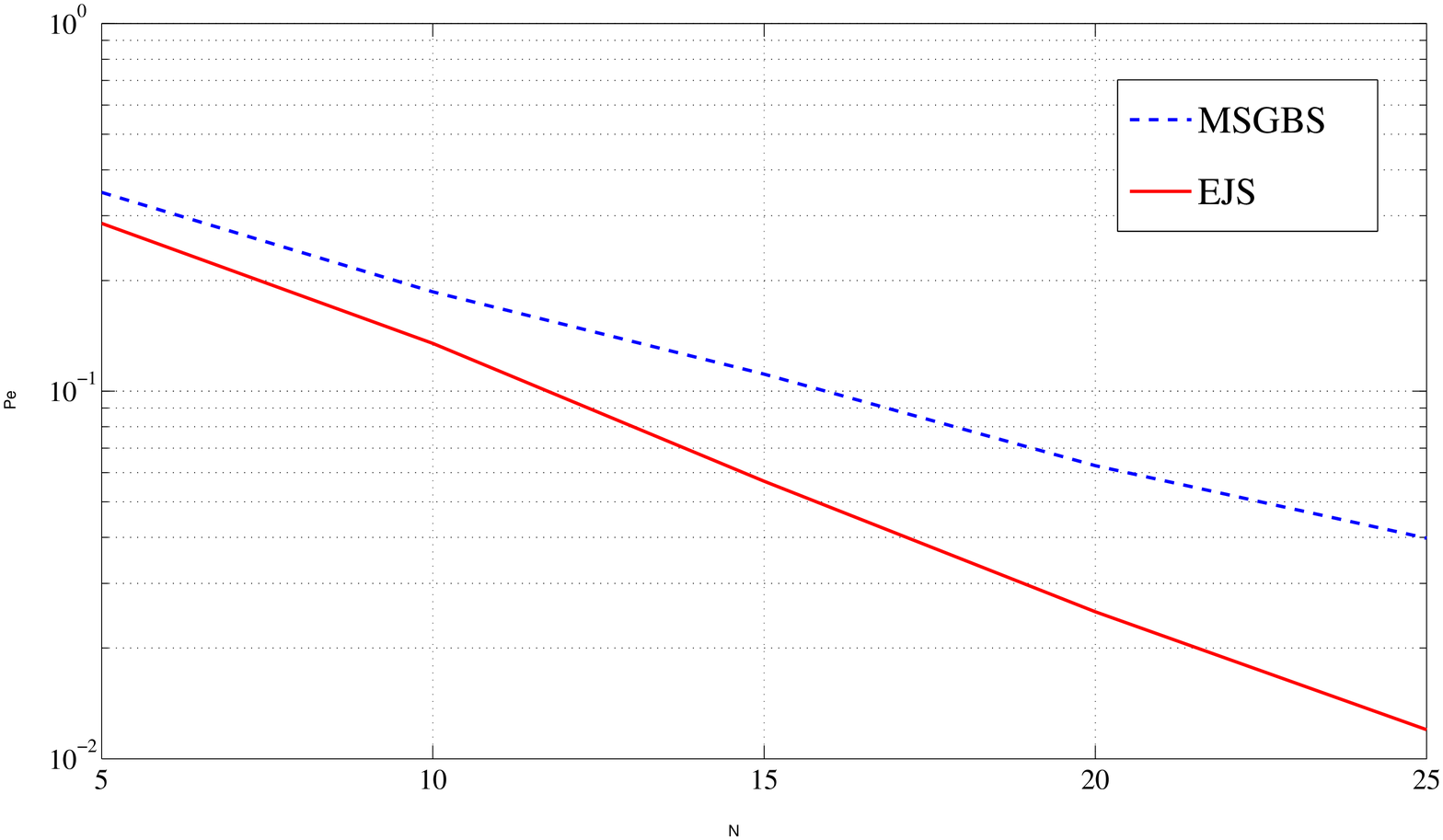}
\caption{Non-sequential noisy generalized binary search with parameter $p=0.2$, total number of label queries $N$, and a rich function class of size $M=5$. The average probability of error is plotted as $N$ varies.}
\label{fig:nonN}
\end{figure}

\begin{figure}[htp]
\centering 
\psfrag{M}{\scriptsize{$M$}}
\psfrag{Pe}{\scriptsize{Pe}}
\psfrag{MSGBS}{\scriptsize{MSGBS}}
\psfrag{EJS}{\scriptsize{$\fc_{EJS}$}}
\includegraphics[width=0.875\textwidth]{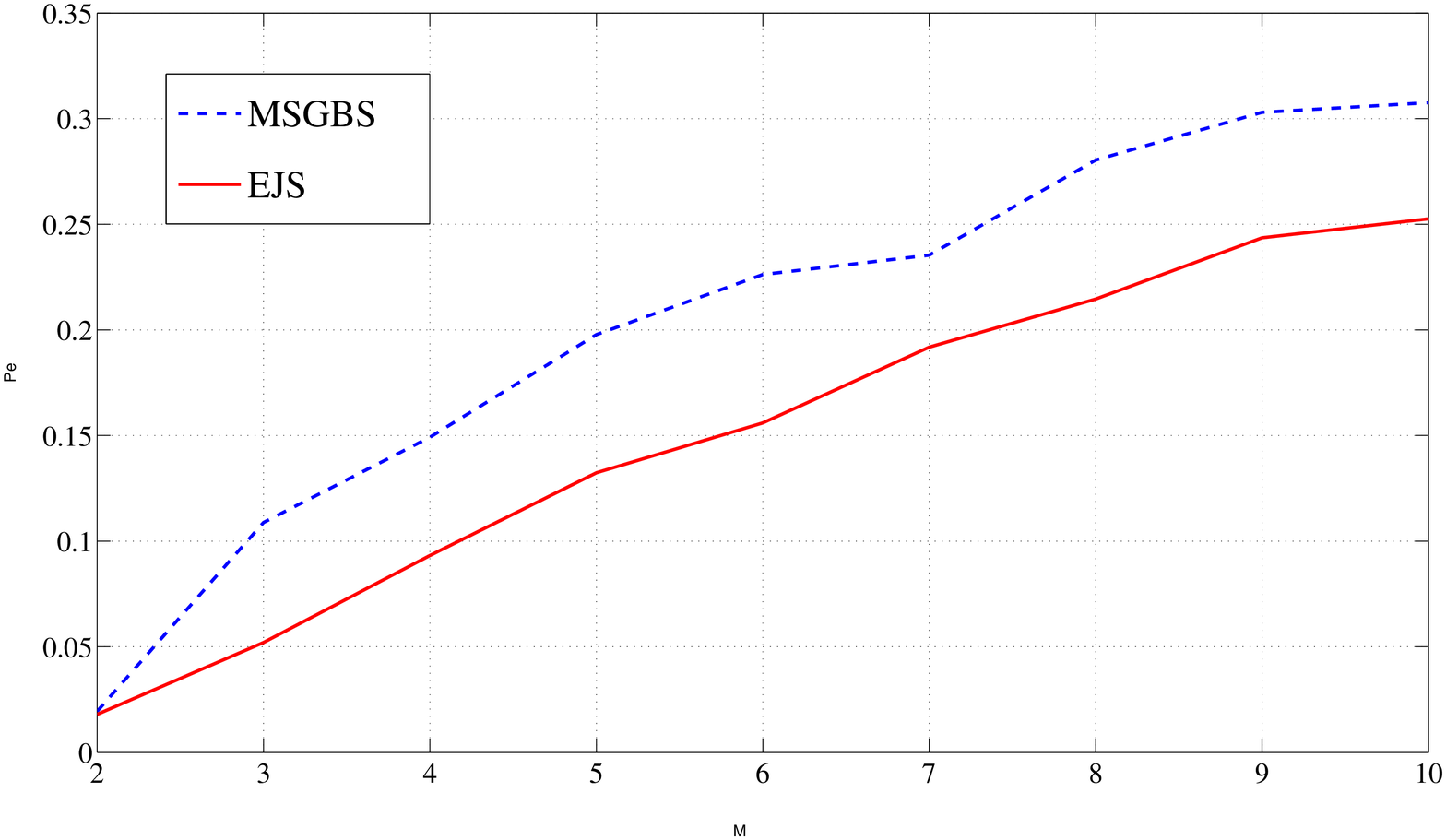}
\caption{Non-sequential noisy generalized binary search with parameter $p=0.2$, total number of label queries $N=10$, and a rich function class of size $M$. The average probability of error is plotted as $M$ varies.}
\label{fig:nonM}
\end{figure}

The figures show the superior performance of $\fc_{EJS}$ over MSGBS in both scenarios and
for different values of $\epsilon$, $N$, and $M$.


\section{Discussion and Future Work}
\label{Discussion}

In this paper, we consider the problem of noisy Bayesian active learning. In this setting, we propose a heuristic policy for querying the labels of samples using Extrinsic Jensen--Shannon divergence, and provide upper bounds on its performance. In addition, we provide information-theoretic lower bounds on the query complexity of any sampling strategy. Comparison to the state-of-the-art~\cite{Nowak11IT} shows that our sampling strategy achieves superior performance for several natural function classes.

Our lower and upper bounds reveal that Bayesian active learning in the presence of noise is a two-phase problem, where the lengths of the phases correspond to the two terms in Theorems~\ref{LB} and~\ref{thm:infoacquisitionUB}. The first phase corresponds to a {\em{search}} among the $M$ functions in the class, and the second phase corresponds to a testing phase where we seek to increase our confidence in the result. An important direction of future research is to extend our algorithms to more general function classes such as linear classifiers and to establish its connection to other notions used to measure the query complexity of active learning such as Alexander's capacity~\cite{DHM07, H07, RR11} and the splitting index~\cite{DAS05}.

\appendices

\section{Proof of Theorem~\ref{LB}}
\label{App:LB}

From Lemma~\ref{lem:TauvsTildeTau}, we have
\begin{align}
\label{tau_vs_tilde_tau_1side}
\mathbb{E} [\tause]
&\ge \mathbb{E} [\tauti] \ (1 - \frac{\epsilon}{\iota}).
\end{align}

Let $V^*_{\iota}:\mathbb{P}(\Omega) \to \mathbb{R}_+$ be the solution to the following fixed point equation: 
\begin{align}
\nonumber
V_{\iota}(\boldsymbol{\rho}) &= \left\{ \begin{array}{ll}
0 & \mbox{if } \max \limits_{j\in\Omega} \rho_{j} \ge 1-\iota \\ 
1 + \min_{x \in \mathcal{X}} \mathbb{E}[V_{\iota}(\boldsymbol{\Phi}^x(\boldsymbol{\rho},Y))], 
& \mbox{otherwise}\end{array}\right.
\end{align}
where ${\boldsymbol{\Phi}}^x$, $x \in \mathcal{X}$, is the Bayes operator defined in~\eqref{PhiDef}. 

It follows from Propositions 9.8 and 9.10 in~\cite{Bertsekas07} that 
\begin{align}
\label{DP_LB}
\mathbb{E}[\tauti] = V^*_{\iota}([1/M, \ldots, 1/M]).
\end{align}

The assertion of the Theorem follows from \eqref{tau_vs_tilde_tau_1side}, \eqref{DP_LB}, and Lemma~\ref{lem:LB_DP} at the end of this section, and by setting $\iota=\frac{\epsilon}{2} \log\frac{4}{\epsilon}$ and $\delta=\frac{1}{\log\frac{4}{\epsilon}}$, as shown below.
\begin{align}
\nonumber
\mathbb{E}[\tause]&\ge 
\left(1-\frac{2}{\log\frac{4}{\epsilon}}\right) \Bigg[ \frac{(1-\frac{1}{\log\frac{4}{\epsilon}}-\frac{\epsilon}{2}\log\frac{4}{\epsilon})\log M -2}{C}+ \frac{\log\frac{1-\frac{\epsilon}{2} \log\frac{4}{\epsilon}}{\frac{\epsilon}{2} \log\frac{4}{\epsilon}} - \log\log\frac{2}{\epsilon} - \log C_2 - 1}{C_1} 
 \Bigg]^+ \nonumber \\
&\ge \Bigg[ \frac{\big(1-\frac{2}{\log\frac{4}{\epsilon}}\big)(1-\frac{\epsilon}{2}\log\frac{4}{\epsilon})\log M -\frac{\log M}{\log\frac{4}{\epsilon}} -2}{C} \nonumber \\
& \quad + \frac{\big(1-\frac{2}{\log\frac{4}{\epsilon}}\big)\log\frac{1}{\frac{\epsilon}{2} \log\frac{4}{\epsilon}} - \log\frac{1}{1-\frac{\epsilon}{2}\log\frac{4}{\epsilon}} - \log\log\frac{2}{\epsilon} - \log C_2 - 1}{C_1} 
 \Bigg]^+ \nonumber \\
&\ge \Bigg[ \frac{\big(1-\frac{2}{\log\frac{4}{\epsilon}}-\frac{\epsilon}{2}\log\frac{1}{\epsilon}\big)\log M -\frac{\log M}{\log\frac{4}{\epsilon}} -2}{C} \nonumber \\
& \quad + \frac{\log\frac{1-\epsilon}{\epsilon} - \log\log\frac{4}{\epsilon} -1 - \log\frac{1-\epsilon}{1-\frac{\epsilon}{2}\log\frac{4}{\epsilon}} - \log\log\frac{2}{\epsilon} - \log C_2 - 1}{C_1} 
 \Bigg]^+ \nonumber \\
&\ge \Bigg[\frac{\big(1-\frac{3}{\log\frac{4}{\epsilon}}-\frac{\epsilon}{2}\log\frac{1}{\epsilon}\big)\log M - 2}{C}+ \frac{\log\frac{1-\epsilon}{\epsilon} - 2\log\log\frac{4}{\epsilon}  - \log C_2 - 4}{C_1}\Bigg]^+. \label{eq:calcul} 
\end{align}

\begin{lemma}
\label{lem:LB_DP}
At any information state $\boldsymbol{\rho} \in\mathbb{P}(\Omega)$ and for any $\iota\in(0,1)$ and $\delta\in(0,1/2)$, 
\begin{align}
\label{lem:LB_DP_eqn}
V^*_{\iota}(\boldsymbol{\rho}) & \ge \left[ \frac{H(\boldsymbol{\rho}) - F_M(\delta) - F_M(\iota)}{C} + \frac{\log\frac{1-\iota}{\iota} - \log\frac{1-\delta}{\delta} - \log C_2 - 1}{C_1} 
{\boldsymbol{1}}_{\{\max \limits_{i\in \Omega} \rho_i \le 1 - \delta \}} \right]^+
\end{align}
where $F_M(z):=H([z,1-z])+z\log (M-1)$ for $0\le z\le 1$.
\end{lemma}

\begin{IEEEproof}
The proof of Lemma~\ref{lem:LB_DP} follows closely the proof of Lemma~1 and Theorem~2 in~\cite{HypJournal} and is provided next.

First we will use the following technical lemma, proved in Appendix~\ref{App:Aux}. 

\begin{lemma}
Any functional $V:\mathbb{P}(\Omega) \to \mathbb{R}_+$ that satisfies the following: 
\begin{align}
\label{ValF_LB}
\displaystyle{
V(\boldsymbol{\rho}) \leq \left\{ \begin{array}{ll}
0 & \mbox{if } \max \limits_{j\in\Omega} \rho_{j} \ge 1-\iota \\ 
1 + \min_{x \in \mathcal{X}} \mathbb{E}[V(\boldsymbol{\Phi}^x(\boldsymbol{\rho},Y))]
& \hspace*{.35in} \mbox{otherwise}\end{array}\right., }
\end{align}
provides a uniform lower bound for the optimal value function $V_\iota^*$. 
\end{lemma}

Next we define $J(\boldsymbol{\rho})=\max \{J'(\boldsymbol{\rho}), J''(\boldsymbol{\rho})\}$ where
\begin{align}
\label{J'Def}
J'(\boldsymbol{\rho})&:=
\left[ \frac{-F_M(\iota)}{C} + \sum_{i=1}^{M} \rho_i \frac{\log\frac{1-\iota}{\iota} - \log\frac{\rho_i}{1-\rho_i} -1}{C_1} \right]^+,
\end{align}
and $J''$ is the right-hand side of \eqref{lem:LB_DP_eqn}, i.e.,
\begin{align*}
J''(\boldsymbol{\rho}) := \left[ \frac{H(\boldsymbol{\rho}) - F_M(\delta) - F_M(\iota)}{C} + \frac{\log\frac{1-\iota}{\iota} - \log\frac{1-\delta}{\delta} - \log C_2 - 1}{C_1} 
{\boldsymbol{1}}_{\{\max \limits_{i\in \Omega} \rho_i \le 1 - \delta \}} \right]^+.
\end{align*}
We show that $J$ satisfies \eqref{ValF_LB} and hence, $V^*_\iota \ge J = \max \{J', J''\} \ge J''$.

We use Jensen's inequality to show that 
\begin{align}
\label{ineqJ'}
J'(\boldsymbol{\rho}) \le 1+\min \limits_{x \in \calX} \mathbb{E}[J'(\boldsymbol{\Phi}^x(\boldsymbol{\rho},Y))], \quad \forall \boldsymbol{\rho}\in\mathbb{P}(\Omega).
\end{align}
For any $\boldsymbol{\rho}$ such that $J'(\boldsymbol{\rho})=0$, inequality \eqref{ineqJ'} holds trivially. 
For any $\boldsymbol{\rho}$ such that $J'(\boldsymbol{\rho})>0$ and for any $x \in \calX$, we have 
\begin{align*}
\mathbb{E}[J'(\boldsymbol{\Phi}^x(\boldsymbol{\rho},Y))] 
&\ge \frac{-F_M(\iota)}{C} + \sum_{i=1}^{M} \int \rho_i f_{h_i(x)}(y) \frac{\log\frac{1-\iota}{\iota} - \log\frac{\rho_i f_{h_i(x)}(y)}{\sum_{j \neq i} \rho_j f_{h_j(x)}(y)} - 1}{C_1} dy \\
&= J'(\boldsymbol{\rho})-
\sum_{i=1}^{M} \rho_i \frac{\int f_{h_i(x)}(y) \log\frac{f_{h_i(x)}(y)}{\sum_{j \neq i} \frac{\rho_j}{1-\rho_i} f_{h_j(x)}(y)} dy}{C_1} \\
&\ge J'(\boldsymbol{\rho})-
\sum_{i=1}^{M} \rho_i \frac{\sum_{j\neq i} \frac{\rho_j}{1-\rho_i} D(f_{h_i(x)}\|f_{h_j(x)})}{C_1} \\
&\ge J'(\boldsymbol{\rho}) - 1.
\end{align*}

For all $\boldsymbol{\rho}$ satisfying $\max \limits_{i\in \Omega} \rho_i > 1 - \delta$,
\begin{align*}
H(\boldsymbol{\rho}) &< (1-\delta)\log\frac{1}{1-\delta} + (M-1) \times \frac{\delta}{M-1}\log\frac{1}{\delta/(M-1)} = F_M(\delta),
\end{align*}
hence, $J''=0$. 
In other words, $J(\boldsymbol{\rho})=J''(\boldsymbol{\rho})>0$ implies that $\max \limits_{i\in \Omega} \rho_i \le 1 - \delta$.

Let $\hat{\boldsymbol{\rho}}=\boldsymbol{\Phi}^x(\boldsymbol{\rho},y)$.
If $\max \limits_{i\in \Omega} \hat{\rho}_i \le 1 - \delta$, then 
\begin{align}
\label{claimJ''a}
J(\hat{\boldsymbol{\rho}}) \ge J''(\hat{\boldsymbol{\rho}}) \ge 
\frac{H(\hat{\boldsymbol{\rho}}) - F_M(\delta) - F_M(\iota)}{C}
+ \frac{\log\frac{1-\iota}\iota - \log\frac{1-\delta}{\delta}-\log C_2 -1}{C_1}. 
\end{align}
On the other hand, if $\max \limits_{i\in \Omega} \hat{\rho}_i > 1 - \delta$, we get
\begin{align}
\label{claimJ''b}
\nonumber
J(\hat{\boldsymbol{\rho}}) &= J'(\hat{\boldsymbol{\rho}})\\ \nonumber
&= \left[ \frac{-F_M(\iota)}{C} + \sum_{i=1}^{M} \hat{\rho}_i \frac{\log\frac{1-\iota}{\iota} - \log\frac{\hat{\rho}_i}{1-\hat{\rho}_i}-1}{C_1} \right]^+ \\ \nonumber
&\stackrel{(a)}{\ge} \left[ \frac{-F_M(\iota)}{C} + \sum_{i=1}^{M} \hat{\rho}_i \frac{\log\frac{1-\iota}\iota - \log\frac{1-\delta}{\delta} - \log C_2 -1}
{C_1} \right]^+ \\
&\ge \frac{-F_M(\iota)}{C} + \frac{\log\frac{1-\iota}\iota - \log\frac{1-\delta}{\delta} - \log C_2 -1}{C_1},
\end{align}
where $(a)$ follows from the fact that under Assumption~\ref{Jump} and for all $i\in\Omega$,
\begin{align*}
\log\frac{\hat{\rho}_i}{1-\hat{\rho}_i} &\le 
\bigg|\log\frac{\hat{\rho}_i}{1-\hat{\rho}_i} - \log\frac{\rho_i}{1-\rho_i} \bigg| + \bigg| \log\frac{\rho_i}{1-\rho_i} \bigg| \\
&\le \bigg|\log\frac{\rho_i f_{h_i(x)}(y)}{\sum_{j\neq i} \rho_j f_{h_j(x)}(y)} - \log\frac{\rho_i}{1-\rho_i} \bigg| + \log\frac{1-\delta}{\delta}\\
&= \bigg|\log\frac{f_{h_i(x)}(y)}{\sum_{j\neq i} \frac{\rho_j}{1-\rho_i} f_{h_j(x)}(y)} \bigg| + \log\frac{1-\delta}{\delta}\\
&\le \log C_2 + \log\frac{1-\delta}{\delta}.
\end{align*}

From the above facts, we obtain: 
\begin{itemize}
	\item {\bf{Case 1:}} For all $\boldsymbol{\rho}$ such that $J(\boldsymbol{\rho})=0$ or $J(\boldsymbol{\rho})=J'(\boldsymbol{\rho})$, it is trivial from \eqref{ineqJ'} that
	\begin{align}
	\label{VLB2case1}
	J(\boldsymbol{\rho}) = J'(\boldsymbol{\rho}) \le 1+\min \limits_{x \in \calX} \mathbb{E}[J'(\boldsymbol{\Phi}^x(\boldsymbol{\rho},Y))] 
	\le 1+\min \limits_{x \in \calX} \mathbb{E}[J(\boldsymbol{\Phi}^x(\boldsymbol{\rho},Y))].
	\end{align}
	
	\item {\bf{Case 2:}} For all $\boldsymbol{\rho}$ such that $J(\boldsymbol{\rho})=J''(\boldsymbol{\rho})>0$,
	and for any $x\in\calX$, we have 
	\begin{align}
	\label{VLB2case2}
	\nonumber
	\mathbb{E}[J(\boldsymbol{\Phi}^x(\boldsymbol{\rho},Y))] &= \int J(\boldsymbol{\Phi}^x(\boldsymbol{\rho},y)) f_x^{\boldsymbol{\rho}}(y) dy \\
	\nonumber
	&\stackrel{(a)}{\ge} \frac{\int H(\boldsymbol{\Phi}^x(\boldsymbol{\rho},y)) f_x^{\boldsymbol{\rho}}(y) dy - F_M(\delta) - F_M(\iota)}{C}\\
	\nonumber
	 & \hspace*{.25in} + \frac{\log\frac{1-\iota}\iota - \log\frac{1-\delta}{\delta}-\log C_2 -1}{C_1} 
{\boldsymbol{1}}_{\{\max \limits_{i\in \Omega} \rho_i \le 1 - \delta \}} \\
	\nonumber
	&= J''(\boldsymbol{\rho}) - \frac{I(\boldsymbol{\rho};f_x^{\boldsymbol{\rho}})}{C} \\
	\nonumber
	&\ge J''(\boldsymbol{\rho}) - 1\\
	&\stackrel{(b)}{=} J(\boldsymbol{\rho}) - 1,
	\end{align} 
	where $(a)$ follows from \eqref{claimJ''a} and \eqref{claimJ''b},  
and $(b)$ holds since $\boldsymbol{\rho}$ is such that $J(\boldsymbol{\rho})=J''(\boldsymbol{\rho})$.

\end{itemize}

Combining~\eqref{VLB2case1} and~\eqref{VLB2case2}, we have that
\begin{align}
	\label{JineqT}
	J(\boldsymbol{\rho}) \le 1+\min \limits_{x \in \calX} \mathbb{E}[J(\boldsymbol{\Phi}^x(\boldsymbol{\rho},Y))].
	\end{align}
What remains is to show that $J(\boldsymbol{\rho}) = 0$ for all $\boldsymbol{\rho}\in\mathbb{P}(\Omega)$ such that $\max_{i\in\Omega} \rho_i \ge 1-\iota$.

For $\boldsymbol{\rho}\in\mathbb{P}(\Omega)$ such that $\max_{i\in\Omega} \rho_i \ge 1-\iota$, we have:
\begin{align}
	\label{J'eq0}
	\nonumber
	J'(\boldsymbol{\rho})&=
\left[ \sum_{i=1}^M \rho_i \frac{\log\frac{1-\iota}{\iota} - \log\frac{\rho_i}{1-\rho_i} -1}{C} - \frac{F_M(\iota)}{C} \right]^+\\
\nonumber
	&\le \left[ \sum_{\{i\in\Omega: \rho_i<1-\iota\}} \rho_i \frac{\log\frac{1}{\iota} + \log\frac{1}{\rho_i}-1}{C_1} - \frac{F_M(\iota)}{C} \right]^+\\
\nonumber
	&\stackrel{(a)}{\le} \left[ \left(\sum_{\{i\in\Omega: \rho_i<1-\iota\}} \rho_i\right) \frac{\log\frac{1}{\iota} + \log\frac{|\{i\in\Omega: \rho_i<1-\iota\}|}{\sum_{\{i\in\Omega: \rho_i<1-\iota\}} \rho_i}-1}{C_1} - \frac{F_M(\iota)}{C} \right]^+\\ \nonumber
	&\stackrel{(b)}{\le} \left[ \frac{\iota\log\frac{1}{\iota} + \iota \log (M-1)}{C_1} - \frac{F_M(\iota)}{C} \right]^+\\
	&\stackrel{(c)}{=} 0,
	\end{align}
	where $(a)$ follows by Jensen's inequality; 
	$(b)$ follows from the facts that $\sum_{\{i\in\Omega: \rho_i<1-\iota\}} \rho_i\le\iota<1$ for any $\boldsymbol{\rho}\in\mathbb{P}(\Omega)$ that satisfies $\max_{i\in\Omega} \rho_i \ge 1-\iota$, and $x \log\frac{1}{x} \le 1$ for $x \in [0,1]$; and $(c)$ holds since $\iota\log\frac{1}{\iota} \le H([\iota,1-\iota])$ and $C\le C_1$. 

On the other hand, for $J''$ and any $\boldsymbol{\rho}\in\mathbb{P}(\Omega)$ such that $\max_{i\in\Omega} \rho_i \ge 1-\iota$, we have:
\begin{align}
\label{J''eq0}
	\nonumber
J''(\boldsymbol{\rho}) &\le \left[ \frac{H(\boldsymbol{\rho}) - F_M(\iota)}{C} + \frac{\log\frac{1-\iota}{\iota} - \log\frac{1-\delta}{\delta}}{C_1} 
{\boldsymbol{1}}_{\{\max \limits_{i\in \Omega} \rho_i \le 1 - \delta \}} \right]^+\\ \nonumber
&\stackrel{(a)}{\le} \left[ \frac{\log\frac{1-\iota}{\iota} - \log\frac{1-\delta}{\delta}}{C_1} 
{\boldsymbol{1}}_{\{\delta\le\iota, \ \max \limits_{i\in \Omega} \rho_i \le 1 - \delta \}} \right]^+\\
&= 0,
\end{align}
where $(a)$ follows from concavity of the entropy function.

Combining~\eqref{J'eq0} and~\eqref{J''eq0}, we have that
\begin{align}
	\label{Jeq0}
	J(\boldsymbol{\rho}) = 0 \quad \text{if } \max_{i\in\Omega} \rho_i \ge 1-\iota.
	\end{align}
It is implied from \eqref{JineqT} and \eqref{Jeq0} that $J$ satisfies \eqref{ValF_LB} and hence, $V^*_\iota \ge J = \max \{J', J''\} \ge J''$. This is a slightly stronger result than~\eqref{lem:LB_DP_eqn}.
\end{IEEEproof}

\section{Proof of Theorem~\ref{thm:infoacquisitionUB}}
\label{App:main}

First let us consider inequality \eqref{UB1stp} in Theorem~\ref{thm:infoacquisitionUB}}.

Notice that for all $i \in \Omega$, upon selecting $X(t)=x$ and observing $Y(t)=y$, the belief state evolves as 
\begin{align*}
\rho_i(t+1)=\rho_i(t) \frac{f_{h_i(x)}(y)}{f_x^{\boldsymbol{\rho}(t)}(y)}. 
\end{align*}

Let $U(\cdot)$ be the average log-likelihood function defined as
\begin{align}
U(\boldsymbol{\rho}) := \sum_{i=1}^{M} \rho_i \log \frac{1-\rho_i}{\rho_i},
\end{align}
and let $\mathcal{F}(t) = \sigma \{X(0), Y(0), \ldots, X(t-1), Y(t-1)\}$
denote the history of samples and observations up to time $t$. 
We have
\begin{align*} 
\lefteqn{\mathbb{E} \left [U(\boldsymbol{\rho}(t+1)) | \mathcal{F}(t) \right ]}\\
&= \sum_{x\in\mathcal{X}} P(X(t)=x) \mathbb{E} \left [\sum_{i=1}^M \rho_i(t+1) \log\frac{1-\rho_i(t+1)}{\rho_i(t+1)} | \mathcal{F}(t), X(t)=x \right ]\\
&= \sum_{x\in\mathcal{X}} P(X(t)=x) \int_{\mathcal{Y}} \sum_{i=1}^M \rho_i(t) f_{h_i(x)}(y) \log\frac{\sum_{j\neq i} \rho_j(t) f_{h_j(x)}(y)}{\rho_i(t) f_{h_i(x)}(y)} dy\\
&= \sum_{i=1}^M \rho_i(t) \log\frac{1-\rho_i(t)}{\rho_i(t)} + \sum_{x\in\mathcal{X}} P(X(t)=x) \sum_{i=1}^M \int_{\mathcal{Y}} \rho_i(t) f_{h_i(x)}(y) \log\frac{\sum_{j\neq i} \frac{\rho_j(t)}{1-\rho_i(t)} f_{h_j(x)}(y)}{f_{h_i(x)}(y)} dy\\
&= U(\boldsymbol{\rho}(t)) - \sum_{x\in\mathcal{X}} P(X(t)=x) \sum_{i=1}^M  \rho_i(t) D(f_{h_i(x)}\|\sum_{j\neq i} \frac{\rho_j(t)}{1-\rho_i(t)} f_{h_j(x)}) \\
&= U(\boldsymbol{\rho}(t)) - \sum_{x\in\mathcal{X}} P(X(t)=x) EJS(\boldsymbol{\rho}(t),x).
\end{align*}

Remember that $\fc_{EJS}$, at any time $t < \tau$, selects a sample that maximizes the EJS divergence, 
i.e., $X(t) = \argmax \limits_{x\in\mathcal{X}} EJS(\boldsymbol{\rho}(t),x)$. 
Thus, under $\fc_{EJS}$, the sequence $\{U(\boldsymbol{\rho}(t))\}$ satisfies 
\begin{align}
\label{alphaIneq}
\nonumber 
\mathbb{E} \left [U(\boldsymbol{\rho}(t+1)) | \mathcal{F}(t) \right ] &= U(\boldsymbol{\rho}(t)) - \max_{x\in\mathcal{X}} EJS(\boldsymbol{\rho}(t),x)\\
&\stackrel{(a)}{\le} U(\boldsymbol{\rho}(t)) - \alpha,
\end{align}
where $(a)$ follows from the assumption of Theorem~\ref{thm:infoacquisitionUB}.
In other words, the sequence $\{-\frac{U(\boldsymbol{\rho}(t))}{\alpha}-t\}$ 
forms a submartingale with respect to the filtration $\{\mathcal{F}(t)\}$.
Let us define a stopping time 
$$\upsilon := \min \left\{ t: \max_{i\in\Omega} \rho_i(t)\ge 1- \min\Big\{\frac{1}{\log 2M},\epsilon\Big\} \right\}.$$ 
It is clear that $\taute \le \upsilon$ and hence, $\mathbb{E}[\taute] \le \mathbb{E}[\upsilon]$ under any query scheme.
By Doob's Stopping Theorem,
\begin{align*}
\frac{-U(\boldsymbol{\rho}(0))}{\alpha} \le \mathbb{E}\left[\frac{-U(\boldsymbol{\rho}(\upsilon))}{\alpha} - \upsilon \right].
\end{align*}
Rearranging the terms, we obtain
\begin{align}
\label{ub:t1c1}
\nonumber
\mathbb{E}[\upsilon]
&\le \frac{U(\boldsymbol{\rho}(0))}{\alpha} + \mathbb{E}\left[\frac{-U(\boldsymbol{\rho}(\upsilon))}{\alpha}\right]\\
\nonumber
&\stackrel{(a)}{\le} \frac{\log M + \mathbb{E}\left[-U(\boldsymbol{\rho}(\upsilon-1))+U(\boldsymbol{\rho}(\upsilon-1))-U(\boldsymbol{\rho}(\upsilon))\right]}{\alpha}\\
\nonumber
&\stackrel{(b)}{\le} \frac{\log M + \max\{\log\log M,\log\frac{1}{\epsilon}\} + \mathbb{E}\left[U(\boldsymbol{\rho}(\upsilon-1))-U(\boldsymbol{\rho}(\upsilon))\right]}{\alpha}\\
\nonumber
&\stackrel{(c)}{\le} \frac{\log M+ \max\{\log\log M,\log\frac{1}{\epsilon}\} +C_2 \left(3 + \frac{1}{\log 2M} \log (M-1)\right)}{\alpha}\\
&\le \frac{\log M+ \max\{\log\log M,\log\frac{1}{\epsilon}\} +4 C_2}{\alpha},
\end{align}
where $(a)$ follows from the fact that initially the functions are equiprobable, i.e., $\boldsymbol{\rho}(0)=[1/M,\ldots,1/M]$ and hence $U(\boldsymbol{\rho}(0))=\log(M-1)$,
$(b)$ holds since $\rho_i(\upsilon-1)<1-\min\big\{\frac{1}{\log 2M},\epsilon\big\}$ for all $i\in\Omega$ and hence,
$$-U(\boldsymbol{\rho}(\upsilon-1))=\sum_{i=1}^M \rho_i(\upsilon-1) \log\frac{\rho_i(\upsilon-1)}{1-\rho_i(\upsilon-1)} < \log\frac{1-\min\{\frac{1}{\log 2M},\epsilon\}}{\min\{\frac{1}{\log 2M},\epsilon\}} < \max\{\log\log M,\log\frac{1}{\epsilon}\},$$ 
and $(c)$ follows from Lemma~\ref{drift:U} in Appendix~\ref{App:Aux}.


The proof of Inequality \eqref{UB2stp} in Theorem~\ref{thm:infoacquisitionUB} follows similar lines. 
Recall from \eqref{rhoTildeDef} that $\tilde{\rho}=1-\frac{1}{1+\max\{\log M,\log\frac{1}{\epsilon}\}}$.
%
Notice that if $\rho_i(t)<\tilde{\rho}$ for all $i \in \Omega$, then
\begin{align*} 
U(\boldsymbol{\rho}(t)) = \sum_{i=1}^M \rho_i(t) \log\frac{1-\rho_i(t)}{\rho_i(t)} > \sum_{i=1}^M \rho_i(t) \log\frac{1-\tilde{\rho}}{\tilde{\rho}} =\log\frac{1-\tilde{\rho}}{\tilde{\rho}}.
\end{align*}

Similar to \eqref{alphaIneq}, we can show that
\begin{align} \label{expctdjump}
\mathbb{E} \left [U(\boldsymbol{\rho}(t+1)) | \mathcal{F}(t) \right ] \le \begin{cases} U(\boldsymbol{\rho}(t)) - \alpha &\mbox{if } U(\boldsymbol{\rho}(t)) > \log\frac{1-\tilde{\rho}}{\tilde{\rho}}\\
 U(\boldsymbol{\rho}(t)) - \beta &\mbox{if } U(\boldsymbol{\rho}(t)) \le \log\frac{1-\tilde{\rho}}{\tilde{\rho}}
 \end{cases} .
\end{align}

Furthermore, from Lemma~\ref{drift:U} in Appendix~\ref{App:Aux}, we know that 
if $\max \limits_{i\in\Omega} \rho_i(t) \ge \tilde{\rho}$, then
\begin{align} \label{bddjump}
\left|U(\boldsymbol{\rho}(t))-U(\boldsymbol{\rho}(t-1))\right| \leq C_2 \left(3 + (1-\tilde{\rho}) \log (M-1)\right) \leq 4 C_2.
\end{align}

The rest of the proof follows directly from (\ref{expctdjump}) and (\ref{bddjump}) and 
Fact~\ref{fact:Martingale} in Appendix IV. 

\section{Noisy Generalized Binary Search}
\label{App:special}

Let $g_p(\cdot)$ and $\bar{g}_p(\cdot)$ be probability density functions on $\mathcal{Y}$ defined as follows:
\begin{align}
\label{gpdef}
g_p(y)=\left\{\begin{array}{ll} p & \mbox{if } y=-1 \\
					       1-p & \mbox{if } y=+1 \end{array} \right. ,
  \ \bar{g}_p(y)= g_p(-y).
\end{align}

It can be easily shown that:
\begin{align*}
C(p)=D(g_p\|\frac{g_p+\bar{g}_p}{2})=D(\bar{g}_p\|\frac{g_p+\bar{g}_p}{2}) \quad \text{and} \quad
C_1(p)=D(g_p\|\bar{g}_p)=D(\bar{g}_p\|g_p).
\end{align*}

%

\subsection{Proof of Proposition~\ref{prop:binary}}
\label{app:propbinary}

The result for the sample-rich class follows from Proposition~\ref{lem:samplerich}.
Next we provide the proof for the class of disjoint interval functions and threshold functions.

\begin{enumerate}
	\item[]
	\item {\bf{Disjoint Class:}}

To prove this case, we will show that
  $$\max_{x\in\mathcal{X}} EJS(\boldsymbol{\rho},x) \ge \max_{i\in\Omega} \rho_i C_1(p).$$
	
	Let $\hat{i}=\argmax \limits_{i\in\Omega} \rho_i$. 
	By definition of the class of disjoint interval functions, there exists a sample $x_{\hat{i}} \in\mathcal{X}$ that satisfies
	$\boldsymbol{h}(x_{\hat{i}})=\boldsymbol{e}_{\hat{i}}$. We have
	\begin{align*}	
	EJS(\boldsymbol{\rho},x_{\hat{i}})
	\ge \rho_{\hat{i}} D\bigg(f_{h_{\hat{i}}(x_{\hat{i}})} \| \sum_{j\neq\hat{i}} \frac{\rho_j}{1-\rho_{\hat{i}}} f_{h_j(x_{\hat{i}})} \bigg) 
	= \rho_{\hat{i}} D(g_p\|\bar{g}_p)
	= \rho_{\hat{i}} C_1(p).
	\end{align*}

%
	
	\item {\bf{Threshold Class:}}
	

We will prove that
  $$\max_{x\in\mathcal{X}} EJS(\boldsymbol{\rho},x) \ge C(p).$$

	At any belief vector $\boldsymbol{\rho} \in \mathbb{P}(\Omega)$, there exists $k$, $k\in\Omega$, such that
	$\sum_{j=1}^{k} \rho_j \le \frac{1}{2}$ and $\sum_{j=1}^{k+1} \rho_j > \frac{1}{2}$. 
	Let $x_{k}$ and $x_{k+1}$ be samples in $\mathcal{X}$ that satisfy
	$\boldsymbol{h}(x_{k})=\boldsymbol{u}_{k}$
	and $\boldsymbol{h}(x_{k+1})=\boldsymbol{u}_{k+1}$ respectively. 
	Let $\delta_1 = \frac{1}{2} - \sum_{j=1}^{k} \rho_j$
	and $\delta_2 = \sum_{j=1}^{k+1} \rho_j - \frac{1}{2}$.
	Notice that $\rho_{k+1}=\delta_1 + \delta_2$.
	There are two cases:
	
\begin{itemize}
	\item Case 1: $\delta_1 \le \delta_2$. We have
	\begin{align*}	
EJS(\boldsymbol{\rho},x_{k}) 
  &= \sum_{i=1}^M \rho_i D\bigg(f_{h_i(x_k)} \| \sum_{j\neq i} \frac{\rho_j}{1-\rho_i}f_{h_j(x_k)} \bigg)\\
  &= \sum_{i=1}^k \rho_i D\bigg(\bar{g}_p \| \frac{1/2 - \delta_1 - \rho_i}{1-\rho_i}\bar{g}_p + \frac{1/2 + \delta_1}{1-\rho_i}g_p \bigg)\\
  &\quad + \rho_{k+1} D\bigg(g_p \| \frac{1/2 - \delta_1}{1-\rho_{k+1}}\bar{g}_p + \frac{1/2 - \delta_2}{1-\rho_{k+1}}g_p \bigg)\\
  &\quad + \sum_{i=k+2}^M \rho_i D\bigg(g_p \| \frac{1/2 - \delta_1}{1-\rho_i}\bar{g}_p + \frac{1/2 + \delta_1-\rho_i}{1-\rho_i}g_p \bigg)\\
	&\stackrel{(a)}{\ge}  (1/2 - \delta_1) D\Big(g_p \| (1/2 + \delta_1) \bar{g}_p + (1/2 - \delta_1) g_p\Big)\\
  &\quad + (\delta_1+\delta_2) D\Big(g_p \| \frac{1}{2} \bar{g}_p + \frac{1}{2} g_p \Big)\\
  &\quad + (1/2 - \delta_2) D\Big(g_p \| (1/2 - \delta_1) \bar{g}_p + (1/2 + \delta_1) g_p \Big)\\
  &\stackrel{(b)}{\ge} D\Big(g_p \| (1-\gamma) \bar{g}_p + \gamma g_p \Big)\\
  &\stackrel{(c)}{\ge} D\Big(g_p \| \frac{1}{2} \bar{g}_p + \frac{1}{2} g_p \Big)\\
	&= C(p),
	\end{align*}
	where 
	$$\gamma = (1/2 - \delta_1)^2 + \frac{1}{2}(\delta_1+\delta_2) + (1/2 - \delta_2)(1/2 + \delta_1),$$
	inequality $(a)$ follows from Fact~\ref{DPQa} in Appendix~\ref{App:Aux} and \eqref{gpdef},
	$(b)$ holds since KL divergence is convex,
	and $(c)$ follows from the fact that
	$\gamma = \frac{1}{2}+\delta_1(\delta_1-\delta_2)  \le \frac{1}{2}$ and by Fact~\ref{DPQa}.

	\item Case 2: $\delta_1 > \delta_2$. We have
		\begin{align*}	
	EJS(\boldsymbol{\rho},x_{k+1}) 
  &= \sum_{i=1}^k \rho_i D\bigg(\bar{g}_p \| \frac{1/2 + \delta_2 - \rho_i}{1-\rho_i}\bar{g}_p + \frac{1/2 - \delta_2}{1-\rho_i}g_p \bigg)\\
  &\quad + \rho_{k+1} D\bigg(\bar{g}_p \| \frac{1/2 - \delta_1}{1-\rho_{k+1}}\bar{g}_p + \frac{1/2 - \delta_2}{1-\rho_{k+1}}g_p \bigg)\\
  &\quad + \sum_{i=k+2}^M \rho_i D\bigg(g_p \| \frac{1/2 + \delta_2}{1-\rho_i}\bar{g}_p + \frac{1/2 - \delta_2-\rho_i}{1-\rho_i}g_p \bigg)\\
	&\stackrel{(a)}{\ge}  (1/2 - \delta_1) D\Big(g_p \| (1/2 - \delta_2) \bar{g}_p + (1/2 + \delta_2) g_p\Big)\\
  &\quad + (\delta_1+\delta_2) D\Big(g_p \| \frac{1}{2} \bar{g}_p + \frac{1}{2} g_p \Big)\\
  &\quad + (1/2 - \delta_2) D\Big(g_p \| (1/2 + \delta_2) \bar{g}_p + (1/2 - \delta_2) g_p \Big)\\
  &\stackrel{(b)}{\ge} D\Big(g_p \| (1-\gamma') \bar{g}_p + \gamma' g_p \Big)\\
  &\stackrel{(c)}{\ge} D\Big(g_p \| \frac{1}{2} \bar{g}_p + \frac{1}{2} g_p \Big)\\
	&= C(p),
	\end{align*}	
	where 
	$$\gamma' = (1/2 - \delta_1)(1/2 + \delta_2) + \frac{1}{2}(\delta_1+\delta_2) + (1/2 - \delta_2)^2,$$
	inequality $(a)$ follows from Fact~\ref{DPQa} in Appendix~\ref{App:Aux} and \eqref{gpdef},
	$(b)$ holds since KL divergence is convex,
	and $(c)$ follows from the fact that
	$\gamma' = \frac{1}{2}+\delta_2(\delta_2-\delta_1) < \frac{1}{2}$ and by Fact~\ref{DPQa}.

\end{itemize}
	
	Therefore,
	\begin{align*}	
	\max_{x\in\mathcal{X}} EJS(\boldsymbol{\rho},x)
	&\ge \max \left\{EJS(\boldsymbol{\rho},x_{k}), EJS(\boldsymbol{\rho},x_{k+1} \right\}\ge C(p).
	\end{align*}


\end{enumerate}

\subsection{Noisy Generalized Binary Search: Asymptotic Analysis}
\label{o1to0}
For disjoint function class $\mathcal{H}_{D}$ and from Theorem~\ref{thm:infoacquisitionUB} and Proposition~\ref{prop:binary},
\begin{align*}
\mathbb{E}[\tause] &\le \frac{\log M + \max\{\log\log M,\log\log\frac{1}{\epsilon}\}}{\frac{1}{M} C_1(p)} + \frac{\log\frac{1}{\epsilon}}{\tilde{\rho} C_1(p)} + \frac{3(4C_2(p))^2}{\frac{1}{M} C_1(p) \tilde{\rho} C_1(p)}\\
&\stackrel{(a)}{\le} \frac{M \log M + M \log\log\frac{M}{\epsilon}}{C_1(p)} + \frac{\log\frac{1}{\epsilon}+1}{C_1(p)} + \frac{6M(4C_2(p))^2}{(C_1(p))^2}\\
&\le \bigg(\frac{M \log M}{C_1(p)}+\frac{\log\frac{1}{\epsilon}}{C_1(p)}\bigg) \times 
\bigg(1+\frac{M \log\log\frac{M}{\epsilon}+1+6M(4 C_2(p))^2/C_1(p)}{M \log M + \log\frac{1}{\epsilon}}\bigg)\\
&= \bigg(\frac{M \log M}{C_1(p)}+\frac{\log\frac{1}{\epsilon}}{C_1(p)}\bigg) (1+o(1)),
\end{align*} 
where $o(1)\to 0$ as $\epsilon\to 0$ or $M\to\infty$ 
and $(a)$ holds since $\frac{1}{\tilde{\rho}}= 1+\frac{1}{\max\{\log M,\log\frac{1}{\epsilon}\}}\le 2$.

For threshold function class $\mathcal{H}_{T}$ and from Theorem~\ref{thm:infoacquisitionUB} and Proposition~\ref{prop:binary},
\begin{align*}
\mathbb{E}[\tause] &\le \frac{\log M+ \max\{\log\log M,\log\frac{1}{\epsilon}\} +4 C_2(p)}{C(p)}\\ 
&\le \bigg(\frac{\log M}{C(p)}+\frac{\log\frac{1}{\epsilon}}{C(p)}\bigg) \times 
\bigg(1+\frac{\log\log M+4 C_2(p)}{\log\frac{M}{\epsilon}}\bigg)\\
&= \bigg(\frac{\log M}{C(p)}+\frac{\log\frac{1}{\epsilon}}{C(p)}\bigg) (1+o(1)),
\end{align*}where $o(1)\to 0$ as $\epsilon\to 0$ or $M\to\infty$.

For rich function class $\mathcal{H}_{R}$ and from Theorem~\ref{thm:infoacquisitionUB} and Proposition~\ref{prop:binary},
\begin{align*}
\mathbb{E}[\tause] &\le \frac{\log M + \max\{\log\log M,\log\log\frac{1}{\epsilon}\}}{C(p)} + \frac{\log\frac{1}{\epsilon}}{\tilde{\rho} C_1(p)} + \frac{3(4C_2(p))^2}{C(p) \tilde{\rho} C_1(p)}\\
&\stackrel{(a)}{\le} \frac{\log M + \log\log\frac{M}{\epsilon}}{C(p)} + \frac{\log\frac{1}{\epsilon}+1}{C_1(p)} + \frac{6(4C_2(p))^2}{C(p) C_1(p)}\\
&\le \bigg(\frac{\log M}{C(p)}+\frac{\log\frac{1}{\epsilon}}{C_1(p)}\bigg) \times 
\bigg(1+\frac{C_1(p) \log\log\frac{M}{\epsilon}+C(p)+6(4 C_2(p))^2}{C(p) \log\frac{M}{\epsilon}}\bigg)\\
&= \bigg(\frac{\log M}{C(p)}+\frac{\log\frac{1}{\epsilon}}{C_1(p)}\bigg) (1+o(1)),
\end{align*} 
where $o(1)\to 0$ as $\epsilon\to 0$ or $M\to\infty$ 
and $(a)$ holds since $\frac{1}{\tilde{\rho}}= 1+\frac{1}{\max\{\log M,\log\frac{1}{\epsilon}\}}\le 2$.

It follows from Proposition~\ref{LB} that
\begin{align}
\label{asymLB}
\nonumber
\mathbb{E}[\tause] &\ge \frac{\log M}{C(p)}\bigg(1-\frac{2}{\log\frac{4}{\epsilon}}-\epsilon\log\frac{1}{\epsilon}\bigg)-\frac{2}{C(p)}+ \frac{\log\frac{1}{\epsilon}}{C_1(p)}\bigg(1-\frac{2\log\log\frac{2}{\epsilon}+\log C_2(p)+4}{\log\frac{1}{\epsilon}}\bigg)\\
\nonumber
&\ge \bigg(\frac{\log M}{C(p)}+\frac{\log\frac{1}{\epsilon}}{C_1(p)}\bigg) \times \bigg(1-\epsilon\log\frac{1}{\epsilon}-\frac{2\log\log\frac{2}{\epsilon}+\log C_2(p)+4+2 C_1(p)/C(p)}{\log\frac{1}{\epsilon}}\bigg)\\
&= \bigg(\frac{\log M}{C(p)}+\frac{\log\frac{1}{\epsilon}}{C_1(p)}\bigg) (1-o(1)),
\end{align} 
where $o(1)\to 0$ as $\epsilon\to 0$.

\section{Technical Lemmas}
\label{App:Aux}

In this appendix, we provide some preliminary lemmas and facts. 
These lemmas are technical and only helpful in proving the main results of the paper.

\setcounter{lemma}{0}

\begin{lemma}
Consider stopping times defined earlier with scalars $\iota \geq \epsilon >0$. We have 
\begin{align*}
\mathbb{E} [\tauti] \ (1 - \frac{\epsilon}{\iota}) \leq \mathbb{E}[\tause] \leq \mathbb{E}[ \taute^*].
\end{align*}
\end{lemma}

\begin{IEEEproof}
Under any query scheme with the stopping rule \eqref{eq:deftau}:
\begin{align*}
\Pe = \mathbb{E} [1-\max_{i \in \Omega} \rho_i(\taute)]\le \epsilon, 
\end{align*}
hence, by construction, 
\begin{align} \label{lb}
\mathbb{E}[\tause] \leq \mathbb{E}[\taute^*].
\end{align}

On the other hand, let us consider $\mathbb{E}[\tauti]$ for any $\iota > \epsilon$. 
Let $\tau_\epsilon$ be a stopping time at which the probability of error satisfies $\Pe\le \epsilon$.
Under any query scheme,
\begin{align}
\nonumber
\mathbb{E} [\taue]
&\ge \mathbb{E}\big[\taue|\max \limits_{j\in\Omega} \rho_j(\taue) \ge 1-\iota\big] \ P\big(\max \limits_{j\in\Omega} \rho_j(\taue) \ge 1-\iota\big) \\
\nonumber
&\stackrel{(a)}{\ge} \mathbb{E}\big[\taue|\max \limits_{j\in\Omega} \rho_j(\taue) \ge 1-\iota\big] \ \Big(1- \iota^{-1} \mathbb{E}\big[1-\max \limits_{j\in\Omega} \rho_j(\taue)\big]\Big)\\
\nonumber
&\stackrel{(b)}{\ge} \mathbb{E}\big[\taue|\max \limits_{j\in\Omega} \rho_j(\taue) \ge 1-\iota\big] \ \big(1 - \frac{\epsilon}{\iota}\big)\\
\label{taueIneq}
&\ge \mathbb{E} [\tauti] \ \big(1 - \frac{\epsilon}{\iota}\big)
\end{align}
where ($a$) follows from Markov inequality 
and ($b$) follows from 
the definition of $\taue$ which implies that $\Pe=\mathbb{E}[1-\max \limits_{j\in\Omega} \rho_j(\taue)] \le \epsilon$.
From~\eqref{taueIneq},
\begin{align}
\mathbb{E} [\tauti] \ (1 - \frac{\epsilon}{\iota}) \leq \mathbb{E}[\tause].
\end{align}
\end{IEEEproof}
 
%

\setcounter{lemma}{2}

\begin{lemma}
Any functional $V:\mathbb{P}(\Omega) \to \mathbb{R}_+$ that satisfies the following: 
\begin{align}
\nonumber
\displaystyle{
V(\boldsymbol{\rho}) \leq \left\{ \begin{array}{ll}
0 & \mbox{if } \max \limits_{j\in\Omega} \rho_{j} \ge 1-\iota \\ 
1 + \min_{x \in \mathcal{X}} \mathbb{E}[V(\boldsymbol{\Phi}^x(\boldsymbol{\rho},Y))]
& \hspace*{.35in} \mbox{otherwise}\end{array}\right., }
\end{align}
provides a uniform lower bound for the optimal value function $V_\iota^*$. 
\end{lemma}

\begin{IEEEproof}
To prove the above fact, we have to slightly modify the state space and introduce
new notations. We assume that after taking the retire-declare action, the system 
goes to the termination state, denoted by $F$, and remains in that state for the rest of the time. 
The state space is modified to $\mathcal{S}=\mathbb{P}(\Omega) \cup \{F\}$ to include the termination state.
For $x \in \calX \cup \{d_1,d_2,\ldots,d_M\}$, $s \in \mathcal{S}$, let
$$c^x(s) =\left\{\begin{array}{ll}
	1 & {{\mbox{if } s=\boldsymbol{\rho}\in\mathbb{P}(\Omega), x\in \calX}} \\
	\infty & {{\mbox{if } s=\boldsymbol{\rho}\in\mathbb{P}(\Omega), \max \limits_{j\in\Omega} \rho_{j} < 1-\iota, x\in\{d_1,\ldots,d_M\}}} \\
	0 & {{\mbox{if } s=\boldsymbol{\rho}\in\mathbb{P}(\Omega), \max \limits_{j\in\Omega} \rho_{j} \ge 1-\iota, x\in\{d_1,\ldots,d_M\}}} \\ 
	0 & {{\mbox{if } s=F}} \end{array}\right. .$$

The Bayes operator is modified as follows:
$$\boldsymbol{\Phi}^x(s,y) =\left\{\begin{array}{ll}
	\boldsymbol{\Phi}^x(\boldsymbol{\rho},y) & {{\mbox{if } s=\boldsymbol{\rho}\in\mathbb{P}(\Omega), x\in \calX}} \\
	F & {{\mbox{if } s=\boldsymbol{\rho}\in\mathbb{P}(\Omega), x\in\{d_1,\ldots,d_M\}}} \\ 
	F & {{\mbox{if } s=F}} \end{array}\right. .$$
Using the notations above, condition \eqref{ValF_LB}
is rewritten as
\begin{align}
\label{Vrewrt}
\nonumber
V(F) &= 0,\\
V(s) &\le \min \limits_{x \in \calX \cup \{d_1,\ldots,d_M\}} \left\{c^x(s) + \mathbb{E}[V(\boldsymbol{\Phi}^x(s,Y))] \right\}, \quad \forall s \in \mathcal{S} - \{F\}. 
\end{align}	

Let $S_0, S_1, S_2, \ldots$ be a sequence of random variables
denoting the belief states at times $t=0,1,2, \ldots$
starting from belief state $s$, i.e., 
\begin{align*}
S_0 &= s,\\
S_{n} &= \boldsymbol{\Phi}^{X(n-1)}(S_{n-1},Y), \quad \forall n, n>0. 
\end{align*}	
Using~\eqref{Vrewrt} iteratively for $N$ times, we obtain
\begin{align*}
V(s) &\le \mathbb{E}_{\pi^*}[c^{X(0)}(s)] + \mathbb{E}_{\pi^*}[V(\boldsymbol{\Phi}^{X(0)}(s,Y))]\\
&=   \mathbb{E}_{\pi^*}[c^{X(0)}(S_0)] + \mathbb{E}_{\pi^*}[V(S_1)]\\
&\le \mathbb{E}_{\pi^*}[\sum_{n=0}^1 c^{X(n)}(S_n)] + \mathbb{E}_{\pi^*}[V(S_2)]\\
&\le \mathbb{E}_{\pi^*}[\sum_{n=0}^{N-1} c^{X(n)}(S_n)] + \mathbb{E}_{\pi^*}[V(S_N)],
\end{align*}	
where subscript $\pi^*$ implies that actions are selected according to an optimal policy $\pi^*$.\footnote{The existence of an optimal policy follows from \cite[Corollary~9.12.1]{Bertsekas07} and since $|\calL|<\infty$.}
Taking the limit as $N \to \infty$, we obtain 
\begin{align*}
V(s) &\stackrel{(a)}{\le} \mathbb{E}_{\pi^*}[\sum_{n=0}^{\infty} c^{X(n)}(S_n)] + \lim \limits_{N \to \infty} \mathbb{E}_{\pi^*}[V(S_N)]\\
&\stackrel{(b)}{=} V^*_\iota(s) + \lim \limits_{N \to \infty} \mathbb{E}_{\pi^*}[V(S_N)]\\
&= V^*_\iota(s) + \lim \limits_{N \to \infty} \mathbb{E}_{\pi^*}[V(F){\bf{1}}_{\{S_N=F\}} + V(S_N){\bf{1}}_{\{S_N\neq F\}}]\\
&= V^*_\iota(s) + \lim \limits_{N \to \infty} \mathbb{E}_{\pi^*}[V(S_N){\bf{1}}_{\{S_N\neq F\}}]\\
&=V^*_\iota(s),
\end{align*}	
where $(a)$ follows from the monotone convergence theorem and $(b)$ follows from the definition of $V^*_\iota$. 
\end{IEEEproof}

\begin{lemma}
\label{drift:log}
For any $i \in \Omega$,
\begin{align*}
\left|\log\frac{\rho_i(t+1)}{1-\rho_i(t+1)}-\log\frac{\rho_i(t)}{1-\rho_i(t)}\right| \leq \log C_2.
\end{align*}
\end{lemma}

\begin{IEEEproof}
\begin{align*}
\left|\log\frac{\rho_i(t+1)}{1-\rho_i(t+1)}-\log\frac{\rho_i(t)}{1-\rho_i(t)}\right|
&= \Bigg|\log\frac{\rho_i(t) f_{h_i(X(t))}(Y(t))}{\sum \limits_{j\neq i} \rho_j(t) f_{h_j(X(t))}(Y(t))}-\log\frac{\rho_i(t)}{1-\rho_i(t)} \Bigg|\\
&= \Bigg|\log\frac{f_{h_i(X(t))}(Y(t))}{\sum \limits_{j\neq i} \frac{\rho_j(t)}{1-\rho_i(t)} f_{h_j(X(t))}(Y(t))}\Bigg|\\
&\le \max_{x \in \mathcal{X}} \sup_{y \in \mathcal{Y}} \log\frac{f_{h_i(x)}(y)}{\min_{j\neq i} f_{h_j(x)}(y)}\\
&\le \log C_2.
\end{align*}
\end{IEEEproof}

\begin{lemma}
\label{drift:rho}
For any $i \in \Omega$,
\begin{align*}
\left|\rho_i(t+1)-\rho_i(t)\right| \leq \rho_i(t)(1-\rho_i(t)) (C_2-1).
\end{align*}
\end{lemma}

\begin{IEEEproof}
\begin{align*}
\left|\rho_i(t+1)-\rho_i(t)\right|
&= \rho_i(t) \left|\frac{f_{h_i(X(t))}(Y(t))}{\sum \limits_{j=1}^M \rho_j(t) f_{h_j(X(t))}(Y(t))} - 1 \right|\\
&= \rho_i(t) \left|\frac{(1-\rho_i(t)) f_{h_i(X(t))}(Y(t)) - \sum \limits_{j\neq i} \rho_j(t) f_{h_j(X(t))}(Y(t))}{\sum \limits_{j=1}^M \rho_j(t) f_{h_j(X(t))}(Y(t))}\right|\\
&= \rho_i(t) (1-\rho_i(t)) \left|\frac{f_{h_i(X(t))}(Y(t)) - \sum \limits_{j\neq i} \frac{\rho_j(t)}{1-\rho_i(t)} f_{h_j(X(t))}(Y(t))}{\rho_i(t) f_{h_i(X(t))}(Y(t)) + (1-\rho_i(t)) \sum \limits_{j\neq i} \frac{\rho_j(t)}{1-\rho_i(t)} f_{h_j(X(t))}(Y(t))}\right|\\
&\le \rho_i(t) (1-\rho_i(t)) \left( \frac{ \max \Big\{f_{h_i(X(t))}(Y(t)),\sum \limits_{j\neq i} \frac{\rho_j(t)}{1-\rho_i(t)} f_{h_j(X(t))}(Y(t))\Big\}}{\min \Big\{f_{h_i(X(t))}(Y(t)),\sum \limits_{j\neq i} \frac{\rho_j(t)}{1-\rho_i(t)} f_{h_j(X(t))}(Y(t))\Big\}}-1\right)\\
&\le \rho_i(t)(1-\rho_i(t)) \left(\max_{k,l\in\mathcal{L}} \sup_{y \in \mathcal{Y}} \frac{f_k(y)}{f_l(y)}-1\right)\\
&= \rho_i(t)(1-\rho_i(t)) (C_2-1).
\end{align*}
\end{IEEEproof}

\begin{lemma}
\label{drift:U}
For any $\delta \in (0,\frac{1}{2}]$, if $\max \limits_{i\in\Omega} \rho_i(t) \ge 1-\delta$, then
\begin{align*}
\left|U(\boldsymbol{\rho}(t))-U(\boldsymbol{\rho}(t-1))\right| \leq C_2 \left(3 + \delta \log (M-1)\right).
\end{align*}
\end{lemma}

\begin{IEEEproof}
Without loss of generality assume $\rho_{\hat{i}}(t)\ge 1-\delta$.
We obtain

\begin{align*} 
\lefteqn{\left|-U(\boldsymbol{\rho}(t-1))+U(\boldsymbol{\rho}(t))\right|}\\
&=\left|\sum_{i=1}^M \rho_i(t-1) \log\frac{\rho_i(t-1)}{1-\rho_i(t-1)} - \sum_{i=1}^M \rho_i(t) \log\frac{\rho_i(t)}{1-\rho_i(t)}\right|\\
&=\left|\sum_{i=1}^M \rho_i(t-1) \left(\log\frac{\rho_i(t-1)}{1-\rho_i(t-1)}-\log\frac{\rho_i(t)}{1-\rho_i(t)}\right) + \sum_{i=1}^M \left(\rho_i(t-1)-\rho_i(t)\right) \log\frac{\rho_i(t)}{1-\rho_i(t)}\right|\\
&\le \max_{i \in \Omega} \left|\log\frac{\rho_i(t-1)}{1-\rho_i(t-1)}-\log\frac{\rho_i(t)}{1-\rho_i(t)}\right| + \left|\sum_{i=1}^M \left(\rho_i(t-1)-\rho_i(t)\right) \log\frac{\rho_i(t)}{1-\rho_i(t)}\right|\\
&\stackrel{(a)}{\le} \log C_2 + \sum_{i=1}^M \left|\rho_i(t-1)-\rho_i(t)\right| \cdot \left|\log\frac{\rho_i(t)}{1-\rho_i(t)}\right|\\
&\stackrel{(b)}{\le} \log C_2 + C_2 \sum_{i=1}^M \rho_i(t)(1-\rho_i(t)) \left|\log\frac{\rho_i(t)}{1-\rho_i(t)}\right|\\
&\le \log C_2 + C_2 \rho_{\hat{i}}(t)(1-\rho_{\hat{i}}(t)) \left|\log\frac{\rho_{\hat{i}}(t)}{1-\rho_{\hat{i}}(t)}\right| + C_2 \sum_{i\neq {\hat{i}}} \rho_i(t) \log\frac{1}{\rho_i(t)}\\
&\stackrel{(c)}{\le} \log C_2 + C_2 + C_2 \bigg(\sum_{i\neq {\hat{i}}} \rho_i(t)\bigg) \log\frac{M-1}{\sum \limits_{i\neq {\hat{i}}} \rho_i(t)}\\
&\le \log C_2 + C_2 + C_2(\delta \log (M-1) + 1)\\
&\stackrel{(d)}{\le} C_2 \left(3 + \delta \log (M-1)\right), 
\end{align*}
where $(a)$ and $(b)$ follow respectively from Lemmas~\ref{drift:log} and~\ref{drift:rho};
and $(c)$ follows from Jensen's inequality and the fact that
$$z(1-z) |\log\frac{z}{1-z}| \leq 1, \ \ \ z \in [0,1];$$
and $(d)$ holds since $C_2\ge 1$ and hence $\log C_2 \le C_2$. 
\end{IEEEproof}

\begin{fact}[Lemma~10 in \cite{EJS-Journal}]
\label{fact:Martingale}
Assume that the sequence $\{\xi(t)\}$,  $t=0,1,2,\ldots$ forms a submartingale with respect to a filtration $\{\mathcal{F}(t)\}$.
Furthermore, assume there exist positive constants $K_1$, $K_2$, and $K_3$ such that
\begin{align*}
& \mathbb{E} [ \xi(t+1) | \mathcal{F}(t) ] \ge \xi(t) +K_1 \hspace{0.1 in} {\mbox{if}} \hspace{0.1 in} \xi(t) < 0,\\   
& \mathbb{E} [ \xi(t+1) | \mathcal{F}(t) ] \ge \xi(t) +K_2 \hspace{0.1 in} {\mbox{if}} \hspace{0.1 in} \xi(t) \ge 0,\\
& \left| \xi(t+1) - \xi(t) \right| \le K_3 \hspace{0.1 in} {\mbox{if}} \hspace{0.1 in} \max\{\xi(t+1),\xi(t)\} \ge 0.
\end{align*}
Consider the stopping time $\upsilon = \min \{ t: \xi(t) \ge B \}$, $B>0$.
Then we have the inequality
$$\mathbb{E} [\upsilon] \le  \frac{B - \xi(0)}{K_2} + \xi(0) {\bf{1}}_{\{\xi(0) <0\}} \left(\frac{1}{K_2} - \frac{1}{K_1} \right)+ 
\frac{3 K_3^2}{K_1 K_2}.$$ 
\end{fact}

\begin{fact}[Lemma~1 in \cite{EJS-Journal}]
\label{DPQa}
For any two distributions $P$ and $Q$ on a set $\mathcal{Y}$ and $\gamma\in[0,1]$, 
$D(P\|\gamma P + (1-{\gamma}) Q)$ is decreasing in $\gamma$.
\end{fact}


}} 

\bibliographystyle{IEEEtran}
\bibliography{HypTest,active1}  


\vspace*{-.8in}
\begin{IEEEbiographynophoto}{Mohammad Naghshvar}
(S'07) received the B.S. degree in electrical engineering from Sharif University of Technology in 2007. 
He obtained the M.Sc. degree and the Ph.D. degree in electrical engineering (communication theory and systems)
both from University of California San Diego in 2009 and 2013, respectively. 
He is currently a senior R\&D engineer at Qualcomm Technologies Inc., San Diego, CA.  
His research interests include active hypothesis testing and optimal experimental design, 
stochastic control and optimization, 
wireless communication and information theory, and routing and scheduling in wireless networks.
\end{IEEEbiographynophoto}

\vspace*{-.9in}
\begin{IEEEbiographynophoto}{Tara Javidi}
(S'96-M'02) studied electrical engineering at the Sharif University of Technology from 1992 to 1996. She received her MS degrees in Electrical Engineering: Systems, and Applied Mathematics: Stochastics, from the University of Michigan, Ann Arbor, MI. She received her PhD 
in electrical engineering and computer science from the University of Michigan, Ann Arbor, in 2002.
From 2002 to 2004, she was an assistant professor at the Electrical Engineering Department, University of Washington, Seattle. In 2005, she joined University of California, San Diego, where she is currently an associate professor of electrical and computer engineering.
%
Her research interests are in communication networks, stochastic resource allocation, stochastic control theory, and wireless communications.
\end{IEEEbiographynophoto}

\vspace*{-.9in}
\begin{IEEEbiographynophoto}{Kamalika Chaudhuri}
is an assistant professor in the Department of Computer Science and Engineering,
University of California, San Diego. She received a bachelor of technology degree in computer science and engineering from the Indian Institute of Technology, Kanpur, in 2002, and a Ph.D. degree in computer science from the University of California at Berkeley in 2007. Her research focuses on the design and analysis of machine-learning algorithms and their applications. In particular, she is interested in privacy-preserving
machine learning, where the goal is to develop machine-learning methods for sensitive data while still preserving the privacy of the individuals in the data set.
\end{IEEEbiographynophoto}

\end{document}